\documentclass[a4paper,usenatbib]{mnras}

\makeatletter

\newcommand{\Rmnum}[1]{\expandafter\@slowromancap\romannumeral #1@}

\newcommand{\gsim}{\lower0.6ex\vbox{\hbox{$\buildrel{\textstyle >}\over{\sim}\ $}}}

\makeatother

\usepackage{newtxtext,newtxmath}
\usepackage{graphicx}	
\usepackage{amsmath}	
\usepackage{subcaption}
\usepackage{booktabs}
\usepackage[dvipsnames,svgnames]{xcolor}
\usepackage{color}

\def\astrid{\texttt{Astrid}}

\def\msd{\, M_{\rm sd}}

\def\kpc{\, {\rm kpc}}
\def\hmpc{h^{-1}{\rm Mpc}}

\def\hkpc{h^{-1}\, {\rm kpc}}

\def\hmsun{{h^{-1} M_{\odot}}}
\def\msun{\, M_{\odot}}

\title[ASTRID Seed IMBHs]{A vast population of wandering and merging IMBHs at cosmic noon}

\author[T. Di Matteo et al.]{
Tiziana Di Matteo$^{1,2}$\thanks{E-mail: tiziana@phys.cmu.edu},
Yueying Ni$^{1,4}$,
Nianyi Chen$^{1}$,
Rupert Croft$^{1,2}$,
Simeon Bird$^{3}$, 
Fabio Pacucci$^{4,5}$ 
\newauthor
Angelo Ricarte$^{4,5}$
Michael Tremmel$^{6}$
\\
$^{1}$ McWilliams Center for Cosmology, Department of Physics, Carnegie Mellon University, Pittsburgh, PA 15213 \\
$^{2}$ NSF AI Planning Institute for Physics of the Future, 
Carnegie Mellon  University, Pittsburgh, PA 15213, USA \\
$^{3}$ Department of Physics \& Astronomy, University of California, Riverside, 900 University Ave., Riverside, CA 92521, USA\\
$^{6}$ Department of Physics, Truman State University, Kirksville, MO 63501, USA \\
$^{4}$Center for Astrophysics $\vert$ Harvard \& Smithsonian, Cambridge, MA 02138, USA\\
$^{5}$Black Hole Initiative, Harvard University,
Cambridge, MA 02138, USA\\
$^{6}$ Department of Astronomy, Yale University, New Haven, CT 06520, USA\\
}

\date{Accepted XXX. Received YYY; in original form ZZZ}

\pubyear{2022}
\begin{document}
\maketitle

\begin{abstract}
Massive black holes in the centers of galaxies today must have grown by several orders of magnitude from  seed black holes formed at early times. Detecting a population of intermediate mass black holes (IMBHs) can provide constraints on these elusive BH seeds. 
Here we use the large volume, cosmological hydrodynamical simulation \astrid, which includes
IMBH seeds and 
dynamical friction to investigate the population of IMBH seeds. Dynamical friction is largely
inefficient at sinking and merging seed IMBHs at high-z. This leads to an extensive population (several hundred per galaxy) of wandering IMBHs in large halos at $z\sim 2$. 
A small fraction of these IMBHs are detectable as 
HLXs, Hyper Luminous X-ray sources. Importantly, at $z\sim 2$, IMBHs mergers produce the peak of GW events. 
We find close to a million GW events in \astrid\, between $z=2-3$ involving seed IMBH mergers.
 These GW events (almost all detectable by LISA) at cosmic noon should provide strong constraints on IMBH seed models and their formation mechanisms. 
At the center of massive galaxies, where the number of IMBHs can be as high as 10-100,
SMBH-IMBH pairs can form. These Intermediate mass ratio inspirals (IMRIs) and extreme mass ratio inspirals (EMRIs), will require the next generation of milli-$\mu$Hz space-based GW interferometers to be detected.
Large populations of IMBHs around massive black holes will probe their environments and MBH causal structure.

\end{abstract}


\section{Introduction}

In the two most recent decades, we have discovered more about supermassive black holes (SMBHs) in the centers of galaxies and their fundamental role in cosmic history than ever before. SMBHs are tiny compared to galactic scales but manifest themselves as quasars and active galactic nuclei fed by gas accretion, which can make them outshine their host galaxy.
Quiescent SMBHs in the center of local galaxies are observed by the way they perturb stellar and gas kinematics in the nuclear regions \citep[e.g.][]{Ferrarese2000, Tremaine2002, Ghez_2008, Genzel_2010}.
Currently, their known local mass spectrum spans values from about $10^{4}-10^{10} \msun$ (see \citealt{Volonteri2021} and references therein for a recent review).
The properties of SMBHs are correlated with those of their host galaxies \citep{Ferrarese2000, Gebhardt_2000} indicating that they play a central role in the dynamics of stars and gas at galactic scales. The relations hint at the importance of powerful AGN outflows which couple the small scales to the larger galaxy scales. 
One key question remains: how do SMBHs actually form? 
SMBHs do not keep the memory of their growth, hence they cannot shed light on their origin. In fact, they likely have grown predominantly via many phases of critical gas accretion and experienced repeated mergers, both of which contribute to erasing any information about an initial BH seed.
The idea that a seed BH has to be formed at early epochs $z>15-20$ is motivated by the discovery of quasars at $z>6$ — now a few hundred are known, with implied BH masses of $10^8-10^9 \msun$, see, e.g., \citealt{Inayoshi2020, Wang2021, Pacucci_2022}.
Thus the SMBH population is thought to arise from a population of seed black holes of ``intermediate'' mass ranging from $\sim 100$ (the remnants of PopIII stars, see, e.g., \citealt{Madau2001, Hirano2014}) to $10^4-10^5 \msun$ (so-called heavy seeds, formed by the direct collapse of gas or dynamical evolution of a massive stellar cluster, see, e.g.,  \citealt{BrommLoeb2003, Inayoshi2020} and the recent review
\citealt{Volonteri2021}). These black holes are often referred to as Intermediate Mass Black Holes (IMBHs).

There is a lot of uncertainty surrounding the correct seeding formation path for SMBHs. No conclusive detections of seed black holes have yet been made, and their number and properties are largely unknown. 
Detecting the seed black hole population close to formation time will remain highly elusive, as they are not expected to be bright enough to be seen at high redshift \citep{Pacucci_2015, Natarajan_2017}.
If seed black holes form binaries they could become sources of gravitational waves (GWs) which could, in principle, be detected by next-generation GW instruments from space \citep{Pacucci_2020}.
However, many independent studies have pointed out that seed BH binaries may not merge effectively at early times due to ineffective  \citep[e.g.][]{Angles2017, Barausse2020, VolonteriHorizon2020, Ma2021, Chen2022}.
The conditions for forming pairs that lead to coalescence in short timescales (i.e., much less than a Hubble time) 
are extremely challenging. BH binaries may form at large separations but require a range of dissipative processes for their orbits to contract. These conditions are unlikely to be fulfilled in the shallow potentials in which seeds originally form.

However, in the cosmic context, galaxies containing seed black holes are embedded in their dark matter halos, which do merge with other halos. 
The halo merger rate evolves over cosmic time (see, e.g., \citep{Dong_2022}), which suggests different frequencies of mergers for major or minor mergers.
In this process, BHs that are  separated by tens to hundreds of kiloparsecs start their orbital evolution down to smaller scales. 
If the mass ratio of merging halos and galaxies is small, as most commonly is the case, the satellite halo is tidally disrupted early in its dynamical evolution. Hence, its central BH is left on a wide orbit too far from the center of the larger galaxy to merge with its SMBH. 
A wandering population of BHs is therefore predicted to exist in galaxies. In some cases, the BHs can pair and eventually merge within a Hubble time.
As a result, we expect the seed population to be in the form of a significant population of wandering IMBHs in galactic halos at later times or possibly found in the center of sub-structures around halos. 
So, perhaps, the most promising tool to distinguish between seeding models comes from finding the black holes that are not in galaxy nuclei (see, e.g., the extensive review by \citealt{Greenreview2020} and references therein).
A key question is how well this population can retain information about the initial mass function imprinted by the seed formation processes at early times. IMBHs that have not yet evolved to SMBH masses represent our best opportunity to understand how they grow and can therefore be used to constrain SMBH formation

Challenging observations of the still elusive population of Hyper-luminous X-ray sources
(HLXs) \citep[e.g.][]{Barrows2019}, and spatially-offset AGN \citep[e.g.][]{Mezcua2020, Reines2020}
are likely electromagnetic counterparts of 
wandering IMBHs (see, e.g., \citealt{Greenreview2020} for a recent review and references therein.)

Because of the ubiquity of SMBHs at lower redshifts, intermediate mass ratio inspirals (IMRIs) of an IMBH with a central SMBH
may also occur \citep{Fragione_2018}, providing clues to the seed formation and properties of the SMBH seeds.
Populations of seed mass black holes in galaxies could give rise to
IMRIs (with mass ratios $q=M_2/M_1 < 10^{-2,-3}$ or extreme-mass ratio inspirals, EMRIs, $q <10^{-4}$). These events are qualitatively 
different from the comparable mass regime of IMBH mergers. The small mass ratio of IMRIs/EMRIs leads to slower evolution of the binary which could reveal the nature of the (dense) stellar environment around galactic nuclei and how SMBHs grow over cosmic time \citep[e.g.][]{barausse2014}. Besides providing information about putative seed black holes, GWs from these IMRIs will carry with them exquisite information about the space-time around the binary and the environment in which they live
\citep[e.g.][]{barack2019}. 

The emergence of a population of IMBHs, its relation to the early seeding population, the onset of black hole mergers, and their relation to their host galaxy mergers are all part of a vastly multi-scale process connected to  galaxy formation, and which involves a rich set of physics.
Cosmological hydrodynamic simulations, which self-consistently model black holes and galaxies, are the methodology of choice to start addressing some of these questions and understand the populations of IMBHs, along with the environments they probe. A number of studies have revealed that a substantial population of BHs in galaxies experience ineffective dynamical friction and do not drift to the center \citep[e.g.][]{Volonteri2016, Tremmel2018a-CHANGA, Tremmel2018b-wanderBH, Pfister2019, Bortolas2020, Bellovary2010, Bellovary2021, Chen2022}. This unavoidably leads to a population of 
AGN spatially offset from their galactic centers, and “wandering” black holes \citep{RicarteIMBH2021, RicarteHLX2021, Weller_2022}. 

Here we extend this previous work to examine the
 IMBH/BH seed population of 'wandering' BHs in 
 the \astrid simulation \citep{Bird2022, Ni2022, Chen2022}. 
The simulation includes a BH seeding model that covers the range of heavy black hole seeds (from $\sim 10^4 \msun$ to $10^5 \msun$ together with a subgrid model for BH dynamical friction. With the large volume and high resolution of \astrid, we can open up the investigation of the IMBH population in the same cosmological simulation setup that follows the growth and emergence of SMBHs.
This has not been possible in cosmological simulations before due to either their limited volume or the lack of subgrid dynamical friction. Several recent cosmological simulations also use BH seeds that are already in the massive black hole range \citep{Tremmel2017, RicarteIMBH2021, Weinberger2018}. \astrid can follow the dynamical evolution of BHs by accounting for unresolved dynamical friction. \cite{Tremmel2018a-CHANGA, Chen2021}, in particular, showed that the formation of massive black hole (MBH) pairs with separations below $\sim 1$ kpc (the precursors to MBH binaries) often occurs after several hundred million years of orbital evolution of the MBH pairs. Moreover, many MBHs do not even form a binary within a Hubble time \citep{Tremmel2018b-wanderBH, RicarteIMBH2021, Chen2022}.


The goal of this paper is to use \astrid to look into the population of IMBHs ("wandering" and merging) at $z\sim 2$ and assess how well they probe the initial seed population, as well as the feasibility of detecting the EM and GW signals associated with them. Note that $z \sim 2$ is close to the peak of the BH merger rates and predicted events for the Laser Interferometer Space Antenna (LISA, see also \citealt{Chen2022}) by {\sc Astrid} (and other simulations such as \citealt{VolonteriHorizon2020}).

In Section \ref{section:asterix-simulation}, we describe the \astrid~simulation. In Section \ref{sec:IMBH_illustration} we describe the IMBH population in \astrid and provide illustrative examples of wandering BHs, seed-seed mergers, and seed-massive black holes IMRIs.
We discuss the occupation fraction and its relation to the seed population and then move on to GW signals from the IMBHs (seed-seed) mergers and IMRIs observable by LISA.
In Section \ref{section:demographics} we then derive the IMBH mass function and occupation fraction, along with their spatial distribution. Moreover, in Section \ref{section:GW_signatures} we describe the GW signatures of the IMBH population in \astrid, before concluding in Section \ref{section6:Conclusion}.

\section[ASTRID simulation]{\astrid\, simulation}
\label{section:asterix-simulation}

\astrid\, is a cosmological hydrodynamical simulation performed using a new version of the MP-Gadget simulation code \citep{Bird2022}. 
It contains $5500^3$ cold dark matter (DM) particles in a $250 \hmpc$ side box, and an initially equal number of SPH hydrodynamic mass elements. 
The initial conditions are set at $z=99$ and the current final redshift is $z=2$. 
The cosmological parameters used are from \citep{Planck}, with $\Omega_0=0.3089$, $\Omega_\Lambda=0.6911$, $\Omega_{\rm b}=0.0486$, $\sigma_8=0.82$, $h=0.6774$, $A_s = 2.142 \times 10^{-9}$, $n_s=0.9667$. 
The mass resolution of \astrid is $M_{\rm DM} = 6.74 \times 10^6 \hmsun$ and $M_{\rm gas} = 1.27 \times 10^6 \hmsun$ in the initial conditions. 
The gravitational softening length is $\epsilon_{\rm g} = 1.5 \hkpc$ for both DM and gas particles.

MP-Gadget (\citealt{Feng:2016,Bird2022}) is related to the Gadget family of cosmological hydrodynamic simulation codes (\citealt{gadget1,gadget2,gadget4}). It is designed  for the exascale era, being highly scalable and  optimized to run on the most massively parallel high-performance computer systems. 
An earlier version of MP-Gadget was used to run the BlueTides simulation~\cite{Feng2015} on the NSF BlueWaters facility, and the most recent version was used for \astrid \citep{Bird2022,NiASTRID2021} on the NSF Frontera supercomputer.

The hydrodynamics, star formation, stellar feedback and patchy reionization models are described in detail
in \cite{Bird2022}. The pressure-entropy formulation of smoothed particle hydrodynamics (pSPH) is used  to solve the Euler equations.
Star formation is implemented based on the multi-phase star formation model of \cite{sh03}, and incorporating several effects following \cite{Vogelsberger:2014}.
A stellar wind feedback model \citep{Okamoto2010} is included, which assumes wind speeds proportional to the local one-dimensional dark matter velocity dispersion. A model is implemented for the return of mass and metal to the interstellar medium from massive stars.
Patchy hydrogen and helium reionization are followed using semi-analytic methods.

\subsection{Black Hole Model}

In \astrid we continue to follow the practice of seeding a BH after the formation of a sufficiently massive halo. We periodically run a Friends-of-Friends (FOF) group finder algorithm and select halos with a total mass \{ $M_{\rm halo,FOF} > M_{\rm halo,thr}$ and stellar mass $M_{\rm *,FOF} > M_{\rm *,thr}$\} to be seeded.
We use $M_{\rm halo,thr} = 5 \times 10^9 \hmsun$ and $M_{\rm *,thr} = 2 \times 10^6 \hmsun$.
The choice of $M_{\rm *,thr}$ is such that BHs are seeded in halos that have cold gas or star formation, as required by most BH seed formation models. Effectively, most of the FOF halos with $M_{\rm halo} > M_{\rm halo,thr}$ already satisfy the $M_{\rm *,thr}$ criteria.

Considering the complex astrophysical process likely to be involved in BH seed formation, we allow haloes with the same mass to have different SMBH seeds.
Therefore, in \astrid instead of applying a uniform seed mass for all the BHs, we probe a range of BH seed mass $\msd$ drawn, in a probabilistic fashion, from the following power-law distribution:
\begin{equation}
\label{equation:power-law}
    P(\msd ) = 
    \begin{cases}
    0 & M_{\rm sd} < M_{\rm sd,min} \\
    \mathcal{N} (M_{\rm sd})^{-n} & M_{\rm sd,min} <= M_{\rm sd} <= M_{\rm sd,max} \\
    0 & M_{\rm sd} > M_{\rm sd,max}
   \end{cases}
\end{equation}
where $\mathcal{N}$ is the normalization factor.
We set $M_{\rm sd,min} = 3 \times 10^4 \hmsun$, $M_{\rm sd,max} = 3 \times 10^5 \hmsun$, and a power-law index $n = -1$ (the power law in the initial mass function for BHs allows us to broadly capture the mass range predicted by a set of heavy seed models, where the more massive seeds are expected to be rarer than less massive ones).
 Our results are therefore applicable to scenarios where seeds have formed with masses $\sim10^{4}-10^{5} \hmsun$, such as through direct collapse or in dense stellar clusters (see, e.g., the reviews by \citealt{Woods_2019, Inayoshi2020}). Much smaller BH seeds formed from, e.g., individual $100 \msun$ stars are not modeled directly here (or in other large volume cosmological simulations), and their evolution, at least at early times, will be different from the BHs in \astrid. Note also, that with our seeding scheme,  any given mass bin (within the range of seed masses above) will have a component due to newly seeded black holes as well as those seeded at earlier stages and have grown from their original seed mass (by accretion/ mergers). In order to distinguish the population of seed BHs to the rest 
 we keep the information of the original seed mass for each BH in \astrid. 
 Finally, our BH seeds are within the range of IMBH and we will refer to them as seed IMBHs if they have masses $M_{\rm BH} < 2 \times M_{\rm seed}$.

For each halo that satisfies the seeding criteria but does not already contain at least one SMBH particle, we convert the densest gas particle into a BH particle. 
Neighboring gas particles are swallowed once (Eddington-limited) accretion has allowed $M_{\rm BH}$ to grow beyond the initial parent particle mass (i.e., once it depletes the gas reservoir of the parent gas particle).

The gas accretion rate onto the BH is estimated via a Bondi-Hoyle-Lyttleton-like prescription \citep{DSH2005}.
We limit the accretion rate to two times the Eddington accretion rate (this has no real implications on any results, compared to imposing the Eddintong limit, \citealt[][]{Ni2022}). 
The BH radiates with a bolometric luminosity $L_{\rm bol}$ proportional to the accretion rate $\dot{M}_\bullet$, with a mass-to-energy conversion efficiency $\eta=0.1$, typical of thin disk accretion \citep{Shakura1973}.
A percentage of 5\% of the radiated energy is coupled to the surrounding gas as the AGN feedback.
The feedback from SMBHs includes what is often referred to as quasar-mode or thermal feedback, as well as kinetic feedback.

The dynamics of the SMBHs are modeled with a newly developed (sub-grid) dynamical friction model \citep{Chen2021} to replace the original implementation that directly repositioned the BHs to the minimum local potential, similar to previous implementations by \citet{Hirschmann2014}, \citet{Tremmel2015}, and \citet{Pfister2019}.
This model provides an improved physical treatment for calculating BH trajectories and velocities,
and with this, BH orbits evolve more realistically, experiencing more gradual orbital decay on timescales that are naturally affected by their mass, initial orbital parameters, and environemt. This is a really important aspect to  capture the dynamics of the BHs in galaxies and  a necessary step as  
cosmological simulation cannot resolve dynamical friction for the BHs as the particle masses are too similar to that of the black holes. 
We expect that when a MBH moves through an extended medium composed of collisionless particles with smaller mass, it experiences a drag force. The source of this force is the gravitational wake from perturbing the particles in the surrounding medium \citep{Chandrasekhar1943} referred to as dynamical friction. 
In addition to the dynamical friction from the collisionless particles (dark matter and stars), we also account for the drag force on the BH from the surrounding gas \citep{Ostriker1999}.

Two BHs merge if their separation is within two times the spatial resolution $2\epsilon_g$, once their kinetic energy is dissipated by dynamical friction and they are gravitationally bound to each other.
The validation of the dynamical friction model in cosmological simulations is described in \cite{Chen2021, Chen2022}.

In \astrid, the minimal BH seed mass is $3\times 10^4 \hmsun$: this is smaller than the stellar and DM particle masses.
Such a small BH mass relative to the surrounding particles causes noisy gravitational forces (dynamical heating) around the BH and thus instability in the BH motion.  Moreover, as shown in some previous works~\citep[e.g.][]{Tremmel2015,Pfister2019}, it is challenging to effectively model dynamical friction in a sub-grid fashion when $M_{\rm BH}/M_{\rm DM} \ll 1$. 
Following \cite{Chen2021}, we introduce another BH mass tracer, the dynamical mass $M_{\rm dyn} \sim M_{\rm DM}$, to account for the force calculation of BH (including the gravitational force and dynamical friction).  
Note that we still use the intrinsic BH mass $M_{\rm BH}$ to account for the BH accretion and AGN feedback. This dynamical mass is much larger than the seed mass of our BHs, which means that dynamical friction would work more efficiently. While this choice certainly influences the dynamics of our black holes, it means that our predictions for the wandering BH population are conservative while still being significantly more accurate than models that force BHs to the centres of galaxies.

A detailed discussion and validation of the BH model (including BH mass, luminosity function, occupation fractions, BH pairs, and mergers) have been presented in \cite{Chen2022, Ni2022, Bird2022}.
 We refer the readers to the first two introductory papers \cite{Bird2022, Ni2022, Chen2022} for a comprehensive description of the simulation code as well as the sub-grid models for star formation and BHs which we only summarize here.

\subsection{Selection of Host Halos and Wandering IMBHs} 
We use the FOF halos and associated subhalos (galaxies) \citep{Ni2022, Bird2022} to identify structure and sub-structure in \astrid. 
IMBHs are seeded at the center of each halo.
Halos will thus start by having a central seed IMBH which, via structure growth and halo mergers, will be incorporated into large halos.
We designate a "wandering" BH to be further than 3 $\times$ the gravitational softening length from the center of the halo. This avoids classifying objects that are about to merge (2 $\times$ the gravitational softening length) as wandering. Some of the wandering BHs are still contained within the same subhalos as the central BH (often referred to as offset nuclear sources). We will use the identification of IMBH within the same galaxy/subhalo in some parts of our analysis.

As BHs are not artificially repositioned at the center of the halo, the distance threshold is used to ensure that the BH is not the central one (simply moving close to the center which is ill-defined within a few gravitational lengths, see \citealt{Chen2021})
and that it is not just a BH about to merge with the central BH.

\subsection{Gravitational Wave Signal from Binary BHs}
BH mergers produce a gravitational signal with a  characteristic strain, $h_s$ as a function of frequency. 
The characteristic strain is given by $h_s(f) = 4f^2 |\tilde{h}(f)|^2$ \citep[e.g.][]{Moore2015}, where $\tilde{h}(f)$ represents the Fourier transform of a time domain signal.
To generate the waveforms, we use the phenomenological waveform PhenomD \citep[][]{Husa2016,Khan2016} implemented within the \texttt{gwsnrcalc} Python package \citep[][]{Katz2019}. 
 The input parameters are the binary BH masses, merging redshift, and the dimensionless spins of the binary. BH masses and redshift in any given merger are part of the simulation output.
 We do not have a model for the dimensionless spin $a$ which characterizes the alignment of the spin angular momentum with the orbital angular momentum, ($a$ ranges from $-1$ to $1$). 
 Consistent with \citet{Katz2020, Chen2021b}, we assume a constant dimensionless spin of $a_1=a_2=0.8$ for all binaries \citep[e.g.][]{Miller2007,Reynolds2013}.


\begin{figure*}
\centering
  \includegraphics[width=1.0\textwidth]{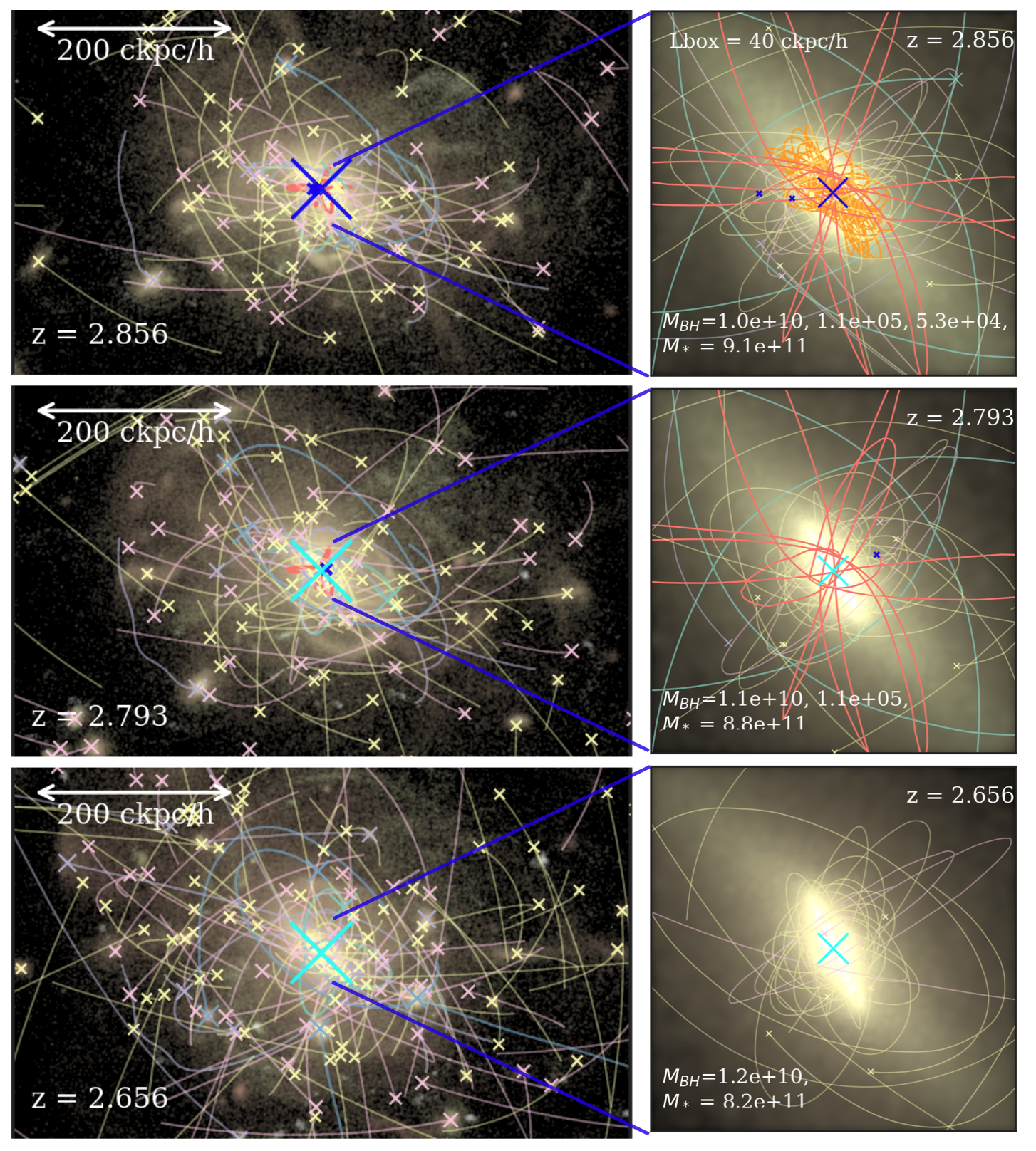}
  \caption{Three snapshots of the evolution (over $\sim$ 200 Myr) of a massive group and its black holes: Group 1 (see Table~\ref{tab:table1} for basic properties). Crosses show the BHs, with the cross size proportional to the mass, over the projected stellar densities.
  The time sequence runs from the top ($z=2.85$) to the bottom ($z=2.65$). 
  \textit{Left panels:} Views of the whole group (over a region of 600 $\hkpc$ on the side), with BHs (from IMBHs through SMBHs) shown as crosses (yellow for seed IMBHs). The tracks for each BH trace its orbital path from $z=3$. The central massive black hole changes colors (dark blue to light blue) when it undergoes its first merger. 
\textit{Right panels:} Views of the central region, the central galaxy, and the local environment of the central black holes. There is a triple BH system, a SMBH in the central galaxy, and two IMBHs orbiting around it (orange and red orbits). These are examples of EMRIs which eventually lead to the mergers of two IMBHs with the central black hole.}
  \label{fig:group9}
\end{figure*}

\begin{table*}
\centering
\begin{tabular}{llllllll}
    \hline
    & group & $M_{\rm halo,FOF}$ [$\msun$] & $M_{*,\rm FOF}$ [$\msun$] & Num BHs & sum $M_{\rm BH}$ [$\msun$] & Num HLXs  & Notes\\
     \hline
    & 1  & $6.0\times10^{13}$ & $1.5\times10^{12}$ & 669 & $2.3\times10^{10}$ & 21 & 2 EMRIs examples\\
    & 2 & $5.0\times10^{13}$ & $1.2\times10^{12}$ & 562 & $8.8\times10^{9}$ & 21 & IMRI example \\
    & 3 & $4.1\times10^{13}$ & $8.3\times10^{11}$ & 495 & $4.4\times10^{9}$ & 23 & 3 IMBH mergers\\
    \hline
\end{tabular}
\caption{Basic properties of the groups shown in Figure~\ref{fig:group9},~\ref{fig:group40},~\ref{fig:group21}. Their halo masses, $M_{\rm halo}$, total stellar masses, $M_{*}$ and black hole mass $M_{\rm BH}$ in units of solar masses and the total number of black holes in each.}
\label{tab:table1}
\end{table*}

\begin{figure*}
\centering
  \includegraphics[width=1.0\textwidth]{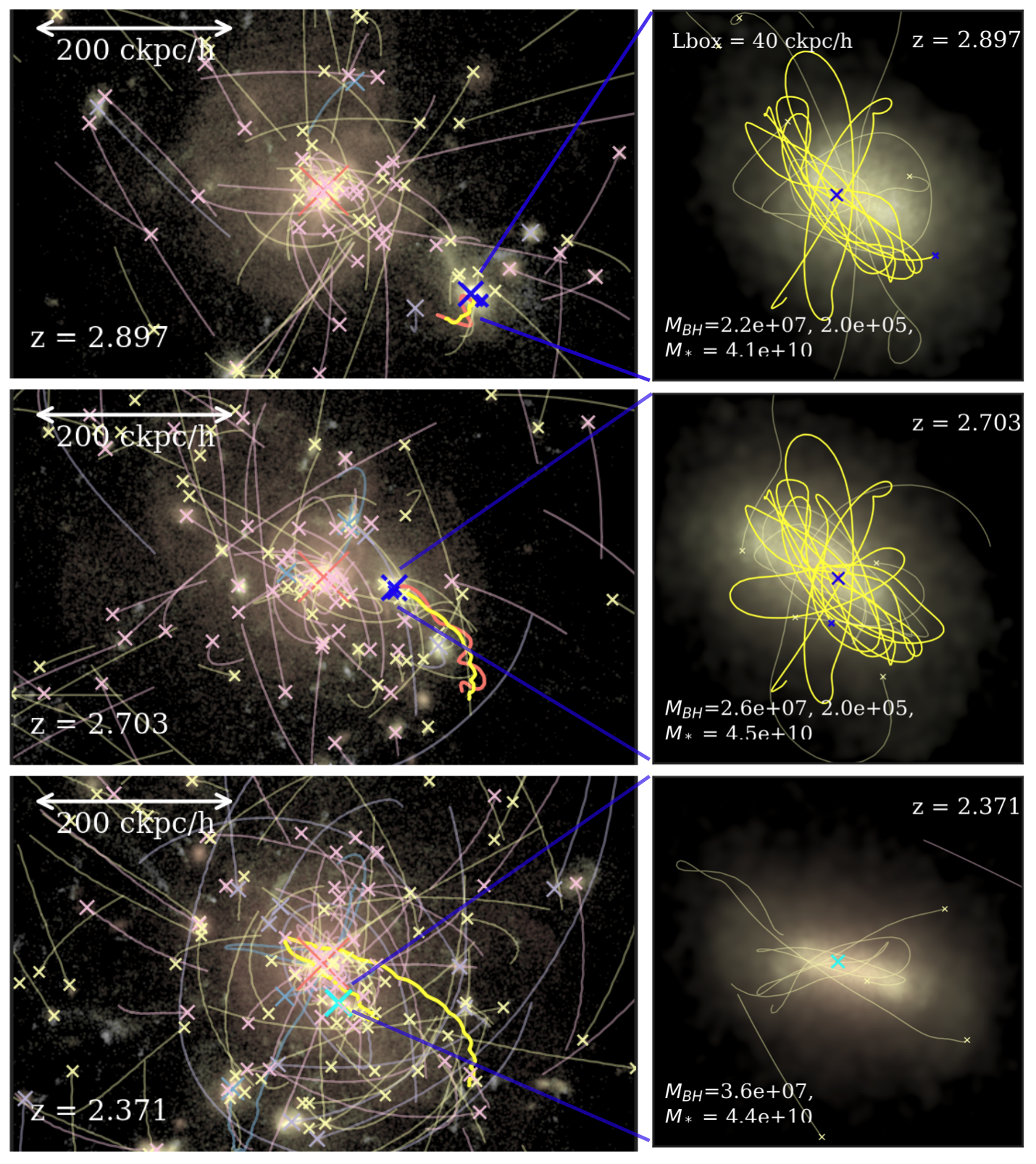}
  \caption{Same as Figure~\ref{fig:group9}, but for Group 2
  (see Table~\ref{tab:table1}). This is another example of a massive group/galaxy with a large population of wandering IMBHs and experiencing a merger with a $M_{*} \sim 10^{10} \msun$ satellite. The sequence from top to bottom spans about $600$ Myrs. There are over five hundred wandering IMBHs. We focus on the infall of a massive satellite with two black holes and IMRIs. 
  The satellite contains a massive $M_{\rm BH} \sim 10^{7} \msun $. As the satellite infalls toward the massive galaxy, its central black hole and an IMBH are also inspiralling. They coalesce before the satellite galaxy merges with the central. This is an IMRI event offset from the center of the main halo/galaxy.}
  \label{fig:group40}
\end{figure*}

\begin{figure*}
\centering
\includegraphics[width=0.95\textwidth]{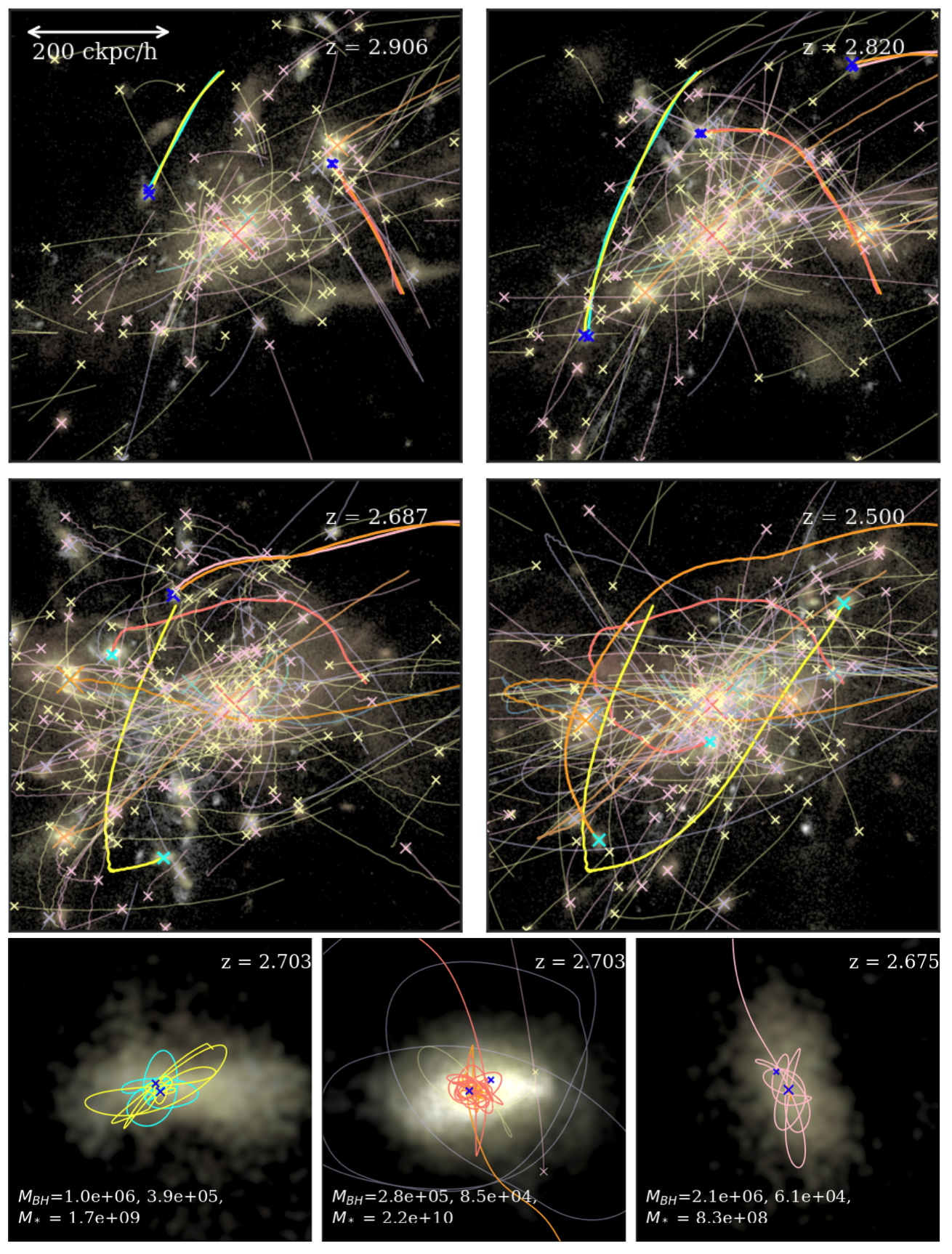}
  \caption{\
  As in Figures~\ref{fig:group9} and ~\ref{fig:group21}, Group 3 is a massive halo (see Table 1) with a large number of wandering IMBH. The time sequence goes from the top left to the bottom right panel (top four panels) showing the projected stellar density, IMBHs and central BH.  The panels in the bottom row show zoom-ins into three separate IMBH-IMBH merger events (in satellite galaxies) shown by the colored tracks in the time sequence.}
  \label{fig:group21}
\end{figure*}

\begin{figure*}
\centering
  \includegraphics[width=1.0\textwidth]{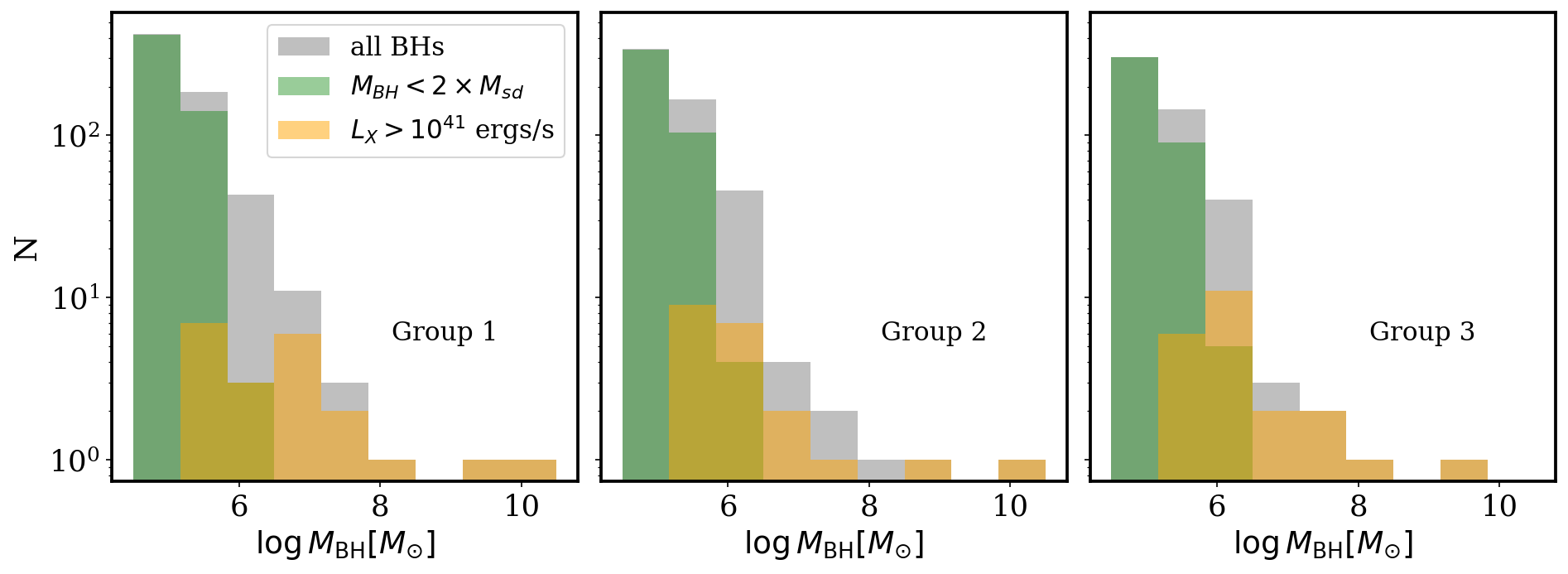}
  \caption{The BH population in the three example galaxy groups, Group 1, 2, and 3 (Figures \ref{fig:group9}-\ref{fig:group21}). The
entire BH population in each group is represented by a grey histogram. The seed IMBH population is displayed in green if it has less than twice the original seed mass. In yellow we display BHs that are HLXs, i.e., those with X-ray luminosity $L_{X} > 10^{41}$ ergs$^{-1}$. A small fraction of IMBHs is HLX sources.}
  \label{fig:BHgroups_hist}
\end{figure*}

\section{The IMBH population in \astrid}
\label{sec:IMBH_illustration}
In our previous paper \citet{Chen2022} we showed that in \astrid, which includes on-the-fly subgrid dynamical friction (DF), the large population of seed black holes does not merge effectively until late times, i.e. $z\sim 3$. This implies delay timescales of order Gyrs to allow the effective sinking of seed BHs.
This conclusion is consistent with a rather large body of recent work 
that has explored the dynamics of BH seeds in idealized galaxies and cosmological simulations, all supporting the conclusion that BH
seeds cannot efficiently sink via DF to galaxy center or be retained unless already massive \citep{Biernacki2017, Tremmel2018b-wanderBH, Pfister2019, Bellovary2019, Barausse2020,Ma2021}.

In this section, we show some examples and basic statistics of the wandering population of seed/IMBHs that result from this ineffective sinking. We look at BHs that end up in substructures or (the larger fraction) which are wandering in  galactic halos.
We concentrate on times close to the peak of the BH merger rates in the simulation, i.e., close to the time when dynamical interactions start becoming more effective. In this way, we can provide examples of some of the 
most common BH merger events.
The large volume and high resolution of \astrid, have not yet allowed us to reach $z=0$. Hence, this study is limited to $z>2$, which is close to the peak of predicted BH merger rates for LISA.

\subsection{Illustrative Examples}
\label{ill}
We start by showing three illustrative examples
of the environments/halos and galaxies at $z\sim 2.5$ which contain 
significant populations of wandering and merging IMBH as well as examples of IMRIs and EMRIs. 
Figure~\ref{fig:group9},  ~\ref{fig:group40}, ~\ref{fig:group21} show, in a time sequence over several hundred million years,
three representative massive groups (see Table~\ref{tab:table1} for their physical properties). The images show the stellar densities and crosses indicate the large population of a few to several 100s IMBHs (Table~\ref{tab:table1}, shown with light-colored crosses, with sizes scaled by mass). 
These three examples show that wandering IMBHs are ubiquitous in \astrid, and occur in large numbers in massive halos. These results are consistent with previous findings from \citet{RicarteIMBH2021, RicarteHLX2021} analysis of the ROMULUS simulation suite.
Amongst the wandering IMBHs, the largest population
is comprised of BHs with mass $< 2 M_{\rm BH, sd}$, i.e. black holes that still retain, approximately, the initial seed mass.

The three sample halo images contain galaxies that span across the full mass range. The total halo virial masses for the three examples are of a few to several $10^{13} \msun$, with central galaxies with masses with $M_{*} \sim 10^{10-11} \msun$.

The distribution of BH masses for each example group is shown in the histogram in Figure~\ref{fig:BHgroups_hist}.
Most of the wandering IMBHs are at the seed mass (also indicated by the light yellow crosses).
At the center  of these groups, there is a massive black hole (indicated by either a red cross or cyan when about to experience a merger or dark blue after a merger event). We also plot the orbits/trajectories for all the BHs in these environments, to show the motion of IMBHs, a large subset of which is orbiting around the center (stripped of their substructure/galaxy in a previous merger), others orbiting around the halos or still attached to their substructures (some paired within a substructure). Finally, some fall in and/or move out as a result of the chaotic dynamical interactions in these crowded systems.

These large-scale environments and the time sequence
we show in these three examples, provide direct visualisation of the backdrop of cosmic structure formation and its relation to the IMBH population. The galaxies, and their dark matter halos, grow through repeated mergers, and that sets the environments in which subsequent evolution occurs. Galaxy (halo) major mergers are relatively rare but minor mergers are frequent. When the mass ratio in the galaxy mergers is small, we see that satellites get tidally disrupted early in their dynamical evolution so that a large population of IMBH/seed BHs are left on a relatively wide orbit. Many of these events have taken place over the evolution of these systems, building a large population of wandering IMBHs.

The histograms (in Fig~\ref{fig:BHgroups_hist})
show the total number of BHs in these three environments. Indeed, the population is dominated by the IMBHs that have retained their initial seed mass (or very close to it). Interestingly, \citet{RicarteIMBH2021} also found consistent results using a different seeding prescription. Most of the seed black holes do not grow effectively in their shallow potentials but instead, end up in large numbers and wandering in larger halos.
The histograms also highlight that the great majority
of the seed black holes are not observable as, e.g., hyper-luminous X-ray sources (HLXs). 

Accreting, wandering black holes, like those shown in
these examples may manifest themselves as HLXs. These are off-nuclear X-ray sources with X-ray luminosities above $10^{41}$ erg/s
\citep{Matsumoto2001, Kaaret2001, Gao2003}. The HLX sources are the most extreme tail of the ultraluminous X-ray (ULX) population, likely to be produced by stellar sources and binaries. HLXs, instead, are expected to be produced by accreting wandering
IMBHs \citep{King2005}. This is also empirically supported by the constraints on their black body temperature obtained from their X-ray spectra \citep{Miller2003, Davis2011}.

Out of the several hundred seed IMBHs (with masses < $2 \times $ their initial seed mass), only a dozen at most are above $L_{x} \sim 10^{41}$ erg/s (when considering Group 1, 2 and 3, see Fig.~\ref{fig:BHgroups_hist}). 
In the next section, we will look at the full population statistics to determine the observability of IMBHs as HLX sources.

We emphasize that the systems we show in these examples are dynamically active, showing evidence of many minor mergers, and are also often associated with some of these minor mergers involving satellite galaxies.

Interestingly, we see several merger events which involve seed IMBHs in these examples.
In the right panels of Figure~\ref{fig:group9},  ~\ref{fig:group40}, ~\ref{fig:group21} we illustrate some of these and the implications for GW signals that may allow us to probe components of this population of seed/IMBHs embedded in these environments.
We emphasise that, using \astrid, we can explore 
structure growth directly, seeing how galaxy halos grow through successive mergers of smaller-scale structures. These structures host IMBHs and central SMBHs with growth approximately consistent with BHs
observed in the local Universe \citep{Ni2022}. The assembly of these SMBHs is accompanied by EMRI (and IMRI) events. We focus
on some examples next.

\subsection{Case I: A couple of EMRIs}
In Figure~\ref{fig:group9}, the number density of IMBHs is so large, and the stellar density in the massive central galaxy so high, that 
the central SMBH ends up paired and, subsequently, merging with
two seed BHs that have indeed sunk to the central regions.

The presence of IMBHs and pairing with the central BH may be ascribed to a number (and probably a combination) of factors. Merger induced torques may helo IMBHs to the central regions.
\citet{Escala2005} found that in hydrodynamic simulations,  clouds of gas,
of gas can induce the decay of orbits 
due to gravitational drag. Also, the effect of chaotic orbits in steep triaxial potentials \citep{PoonMerrit2004}(which likely result from repeated mergers in these galaxies at these redshifts)  
can lead to the extraction of energy and angular momentum.
Finally, 
it has also been suggested \citep{HoffmanLoeb2007} that a third BH closely interacting may induce
 perturbation in the 
potential and possibly shorten merger timescales.

In the inset, we zoom into the central galaxy and show the evolution (orange and red orbits) for these two consecutive  merger events. The first EMRI (mass ratio $q=10^{-5}$), involves a seed BH ($M_{\rm BH} = 5\times10^{4} \msun$ with the central
$\sim 10^{10} \msun$ SMBH. As shown by the orange orbits, the seed BH has paired and has been orbiting the central massive object for a long time. 
The coupling with a second seed BH (shown with the red tracks) on a wide eccentric orbit, likely allows the first merger to finally take place.
This first merger is quickly followed by a second with another seed/IMBH of $M\sim 10^5 \msun$. These are uncommon events
in the simulation, which would lead to an EMRI involving a SMBH and an IMBH. Also, we note that events involving EMRIs with a $10^{10} \msun$ SMBHs would not be observable in the LISA frequency range (see also Sec. \ref{section:GW_signatures}) but would require next-generation GW facilities (e.g. AMIGO, or $\mu$Ares \citealt{amigo20, Sesana_2021}). 

These events are extremely interesting: the small mass ratio of
EMRIs lead to a slow evolution of the binary. As such, the light secondary completes many orbits (even at the scales of the simulations this is evident). We expect such a binary (at much smaller scales) to orbit at relativistic speeds in the strong gravitational field of the heavy primary. The GWs carry with them exquisite information about the space-time of the binary and the environment surrounding the EMRI.

\subsection{Case II: An IMRI}
Figure~\ref{fig:group40} shows another relatively massive group with a central SMBH of $10^{9} \msun$ in
a galaxy of $M_{*} \sim 10^{12} \msun$, surrounded by a large number of satellites and substructure. As in the previous example,
the population of IMBHs consists of several hundred objects.
Interestingly, there is a relatively massive satellite
($M_{*} \sim 10^{10} \msun$) infalling toward the center of the massive group and shown by the four subsequent frames. This contains two paired BHs that end up merging. The evolution of the BHs in this system as it infalls is shown in the zoomed region. 
For this event, the massive black hole is $2-3 \times 10^{7} \msun$ and paired with a IMBH of $2\times 10^{5} \msun$. As the galaxy is infalling, the two black holes orbit each other. As the system evolves, for much of the time even the primary is often off-centered and the secondary, seed BHs is on a wide orbit, often displacing it in the outskirts of the galaxy/halo.
The primary is detectable as an AGN: its bolometric luminosity is typically above $10^{44}$ erg/s but the secondary will remain undetectable throughout.
This $q=M_2/M_1=0.01$ event is an example of the IMRI, which would be detectable by LISA.

\subsection{Case III: Equal mass mergers with seed IMBHs}
In this final example (the zoomed-in regions in Figure~\ref{fig:group21}) we show a set of dwarf satellites which host seed/IMBHs as they orbit the central galaxy and experience mergers. 
These are seed/IMBHs in substructures that end up paired (as a result of a previous merger) and eventually merging. These events constitute the most common type of pairing, and the majority of merger events \citep{Chen2022} involve two seed black holes in dwarf galaxies ($M_* < 10^9 \msun$. 
Seed BHs fail to efficiently sink to and are trapped in the galactic center via dynamical friction at high-z \citep{Ni2022, Ma2021}. 
The pairs eventually form but do not merge until much later (as in this example, below $z \sim 3$ in the simulations). As discussed in \citet{Ni2022} the merger rate is dominated by these types of events and peaks between $z=2-3$. These are the most commonly observable events for LISA. 

\section{IMBH Demographics}
\label{section:demographics}

\subsection{IMBH Mass function}
\label{section3:BH-statistics}
In \astrid~we can inspect the mass function of IMBHs ($M_{\rm BH} \lesssim 10^6 \msun$).
IMBHs include those residing at the centers of galaxies and those wandering, or off-center. A fraction of the IMBH population is composed of seed BHs (at $z\sim 2$).

\begin{figure}
\centering
  \includegraphics[width=1.0\columnwidth]{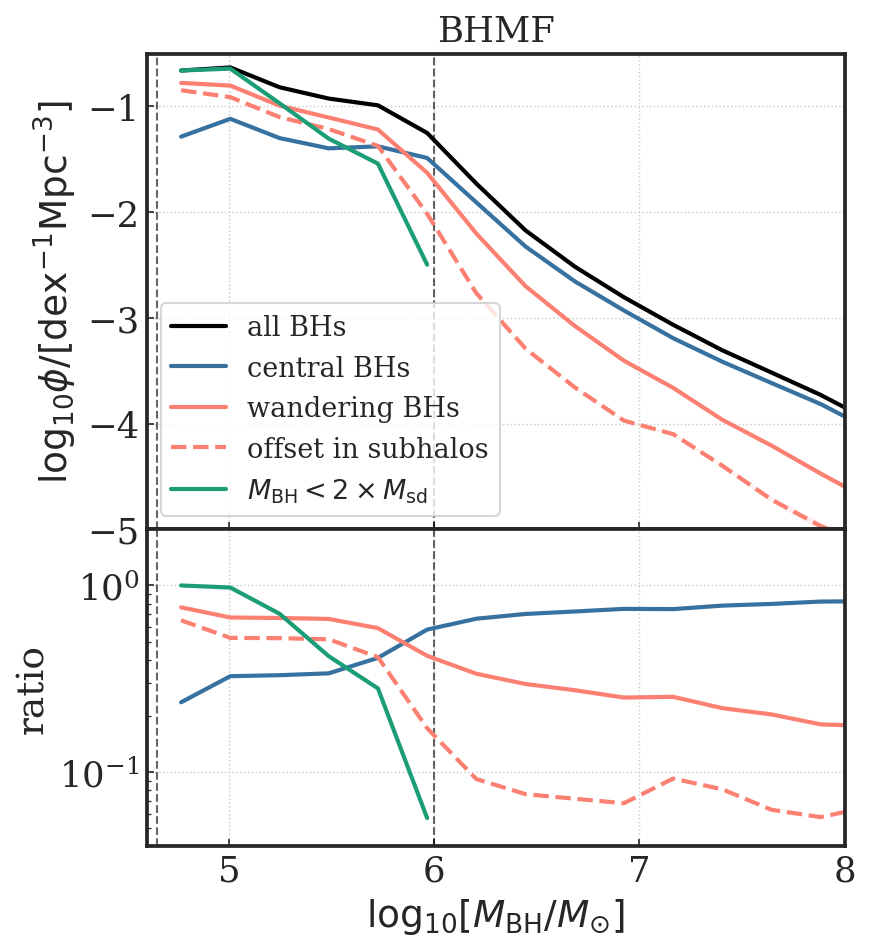}
  \caption{The total BH mass function at z=2. is shown in black. IMBHs ($< 10^6 \msun$) are dominated by seed BHs (green line) and wandering BH (solid orange line). The dashed line shows the component of wandering BHs that reside within a subhalo rather than the large halo. The largest component of the massive black holes is central (the component of massive wandering BHs is due to major mergers).}
  \label{fig:BHmassfn}
\end{figure}

In Figure~\ref{fig:BHmassfn}, we plot the total BH mass function $z\sim 2$ in \astrid~from the minimum (seed) BH mass, up to $10^8 \msun$. We display the contributions due to (a) central, and (b) (off-center) wandering, further divided into (i) wandering black holes, and (ii) off-set nuclear sources (i.e., those wandering but still contained in the same subhalo/galaxy  but are not the central BH). Finally, we show directly the seed BH component (d).

Seed black holes make up the great majority of IMBHs at $M_{\rm BH} \lesssim 10^{5} \msun$. This fraction decreases to less than 1 in 10 at $M_{\rm BH}\sim 10^6 \msun$. The mass function at the low
mass flattens out, thus encoding the information of our seeding procedure (a power law seed mass function). The mass function is significantly more flat below $M_{\rm BH}\sim 10^6 \msun$, the upper limit for our seed black holes.

Consistent with \citet{RicarteIMBH2021}, about 80\% of IMBHs are 'wandering', 60\% of IMBHs are off-center within the same galaxy (or substructure), what we refer to as offset in subhalos in Figure~\ref{fig:BHmassfn}. The remaining fraction of the wandering population is spread over the halo (and/or in a separate galaxy or substructure).

Above black hole masses $M_{BH} \sim 10^6\msun$, the contribution from central black holes dominates the mass function. At these large BH masses, only 1 in 10 massive black holes is offset from the nuclear BH, or 'wandering'. The transition occurs at masses $M_{BH} \sim 10^6 \msun$, where approximately half of BHs are wandering and half are central.

\subsection{IMBH Occupation fraction}
\begin{figure*}
\centering
   \includegraphics[width=1.0\textwidth]{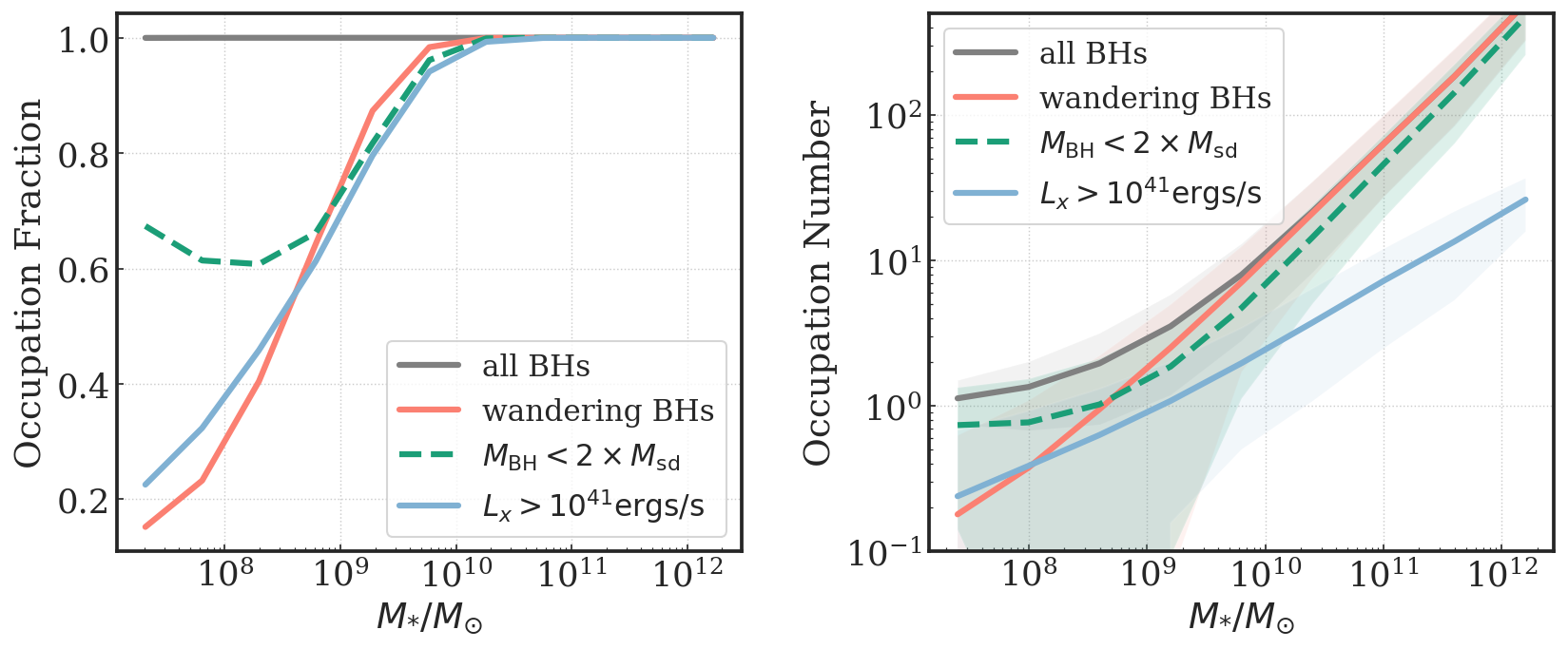}
  \caption{All BHs are shown in grey, the wandering IMBHs in orange, the seed IMBHs with green dashed lines, and the HLX component in blue.
  \textit{Left panel:} 
  The occupation fraction is shown as a function of stellar mass of the central galaxy in the halo. Occupation fractions for off-center, wandering black holes reaches $60\%$ at $M_{*} \sim 10^{9} \msun$. 
  \textit{Right panel:} The BH occupation numbers as a function of total stellar mass. Wandering black holes number in the several hundred in large galaxies. 
  }
  \label{fig:occupfrac}
\end{figure*}

In Figure~\ref{fig:occupfrac} we show the occupation fractions of IMBHs as a function of the host galaxy's stellar mass.

As expected, in the simulations the occupation fraction for BHs
approaches $1$ for all galaxies. This suggests that the vast majority of galaxies contain an IMBH /seed or SMBH. The occupation fraction of
the seed IMBHs is about $60\%$ up to $10^9 M_{*}$, (the rest of BHs have grown above this mass). Seed mass IMBHs are found in $100\%$ of galaxies with $M_{*}  > $ a few $10^{10} \msun$  and constitute the largest contribution to the wandering BH population. Similar results were also found by \citet{Sharma2022} for ROMULUS at $z\sim 2$.
 
The occupation fraction of wandering black holes drops
to about 50\% at $M_* \sim 10^{9} \msun$ and tracks closely the seed BH occupation fraction 
down to these stellar mass.
(The seed population remains around 60\% below this stellar mass). 
For  $M_* \sim 10^8 \msun$, 
the wandering occupation fraction drops down to 20\% implying that most systems will not have any significant population of IMBHs
(beyond the central one) at low masses.

As in \citet{Ni2022}, we calculate the X-ray band luminosity is estimated by converting AGN $L_{\rm Bol}= \eta \dot{M}_{\mathrm{BH}} c^2$ (for $\eta=0.1$, the efficiency of a radiatively efficient accretion disk) to the hard X-ray band [2-10]~keV following the luminosity dependent bolometric correction $L_X = L_{\rm Bol}/k$ from \cite{Hopkins2007}.
The occupation fraction of accreting IMBHs (able to produce an HLX ($L_{x} > 10^{41} $erg/s) is also close to $20\%$,
for the $M_{*} \sim 10^{8} \msun$. 
This is likely contributed by IMBHs growing beyond their original seed mass (see, e.g., \citealt{Pacucci_2021_active} who also derives a similar, mass-dependent, percentage for the active population of BHs).
The occupation fraction for HLXs approaches  $50\%$ at 
$M_* \sim 10^{9} \msun $, tracking closely (albeit slightly below) that of wandering IMBHs.
In summary, about one in two galaxies with $M_* \sim 10^{9} \msun $ has at least one offset, wandering IMBH, close to the seed mass. The occupation fraction of wandering BHs grows to 100\% at $M_* > 10^{10} \msun$.

In the right panel of Fig.~\ref{fig:occupfrac}, we show the actual mean occupation number of IMBHs as a function of stellar mass.
For $M_* \sim 10^{10} \msun$ there are typically 10 wandering IMBHs and this population grows to several hundred at $M_* \sim 10^{12} \msun$. Seed IMBHs are indeed the largest component of the 
wandering BH population. This is supported by the examples shown in Section \ref{sec:IMBH_illustration}. 
With several hundreds IMBHs, the mean number of HLXs is expected to be a few tens even for the largest galaxies. Typical HLX numbers become close to unity at $M_{*} \lesssim 10^{10} \msun$ (with a significant fraction of those being close to the seed IMBHs). 
The high-mass stellar systems tend to have a small fraction of accreting IMBHs, close to 0.1, implying that roughly 90\%
 of the IMBH population in these large systems does not accrete effectively and remains largely ``invisible.'' This decrease in the fraction of the luminous wandering BH as a function of stellar mass is also seen in the Romulus simulation \citep{RicarteIMBH2021}.


\subsection{Spatial Distribution of IMBHs}

\begin{figure}
\centering
\includegraphics[width=1\columnwidth]{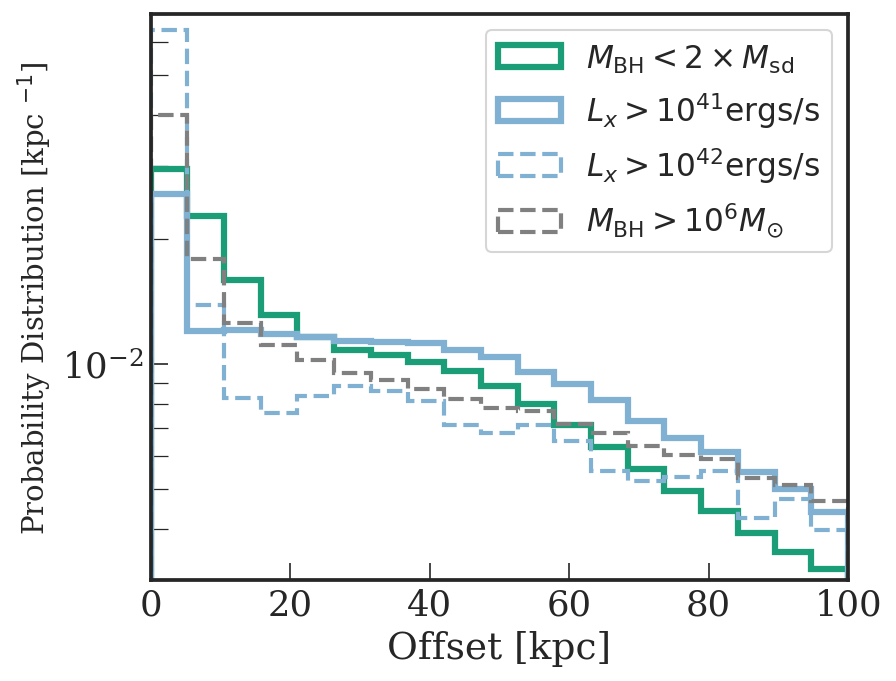}
  \caption{Probability distribution of black hole population (green: IMBHs; blue: HLXs as a function of the offset distance to the galaxy center). There is a broad distribution of IMBHs but also a significant peak for the seed IMBHs within 20 kpc. The associated HLX sources show a flat distribution between 20-40 kpc.
  }
  \label{fig:offsets1}
\end{figure}

In Figure~\ref{fig:offsets1} we plot the normalised spatial distribution of IMBHs as a function of radial distance from the center of the main galaxy/halo in which they reside. This distance is referred to as the offset. 
The peak of the distribution is at small offset values: these are the central black holes in galaxies (which we include here). The wide part of the distribution and tail encompass the wandering IMBHs component
with a broad and rather flat peak for the HLX population at around 20-40 kpc (recall HLXs also include MBHs). While IMBHs are predominant at the center of galaxies, their spatial distribution spans the full radial extent of halos. 

The wandering seed IMBHs have a similar distribution, albeit somewhat steeper than the HLX sources at all scales. Seed IMBHs
accumulate within the central $20$ kpc, but are mostly stripped of their host galaxies and, hence, do not accrete effectively \citep[see also][]{Chen2022, RicarteIMBH2021}. Interestingly
the distribution of HLXs traces that of massive black holes ($M_{\rm BH} > 10^6 \msun$) for offsets $> 60$ \kpc\  (likely enhanced by BHs in early stages of infall and the frequent galaxy mergers 
at $z\sim 2$).
For completeness, we also show the total distribution for $L_{\rm x} > 10^{42}$ erg/s.
Aside from the innermost radial bin (i.e., the central black holes) the distribution traces that of massive black holes.
This demonstrates even more clearly that IMBHs are unlikely contributors of HLX sources at  $L_{\rm x} > 10^{42}$. We will discuss this in the next section as we compare simulations to observed HLX populations.

In Figure~\ref{fig:HLXmap} we show six example regions centered in moderate mass galaxies ($M_* \sim 10^9-10^{10} \msun$) with HLX and IMBHs shown in red and white crosses, respectively. These examples show that many of these galaxies are perturbed, and undergoing significant mergers. Often for regions out to several tens of kpc, there are multiple subhalos and substructures containing HLX sources out to these distances. This explains the relatively flat distributions shown in Figure~\ref{fig:offsets1}.

In summary, \astrid predicts a significant population of IMBHs within a few tens of kpc from central galaxies. These are however inefficient accretors and unlikely to be HLX sources. The broad distribution of HLX sources implies that at least a fraction of HLX observed in the outer regions of galaxies or substructures (offsets of $>30-60$ kpc) can be explained by the presence of IMBHs.

\subsubsection{Comparison with 
Hyperluminous X-ray Sources}
\begin{figure*}
\centering
  \includegraphics[width=1.0\textwidth]{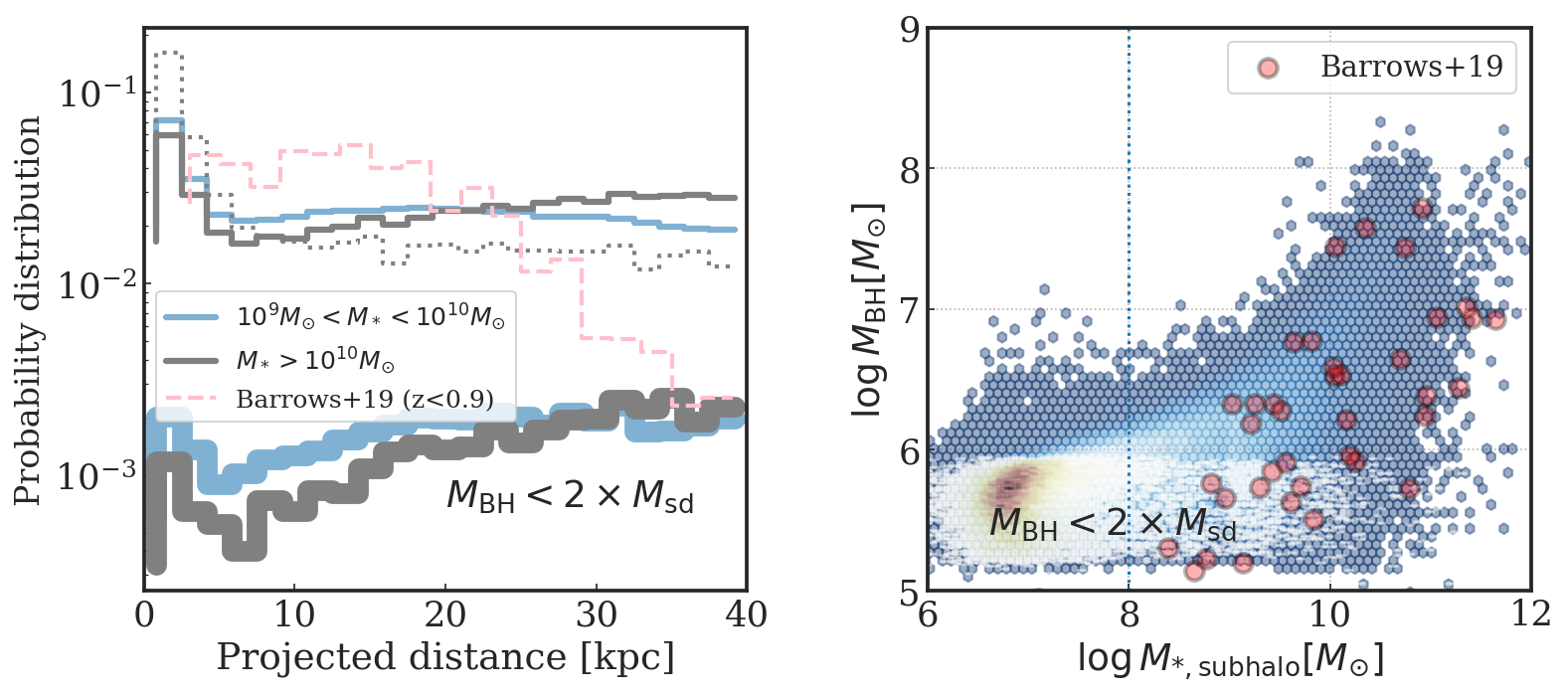}
  \caption{
  \textit{Left panel:} 
  The probability distribution of ULX sources as a function of projected distance from the galaxy center. The distribution for two stellar mass bins is shown: light blue for $10^9 \msun < M_* < 10^{10} \msun$ and grey for $M_* > 10^{10} \msun$. The thick lines at the bottom show the fractional contribution to the ULX distribution from seed IMBHs.
  \textit{Right panel:} The $M_{\rm BH} - M_*$ relation for wandering ULX sources.
}
  \label{fig:ULXoffsets}
\end{figure*}

In Figure~\ref{fig:ULXoffsets} 
we show the accreting, wandering black holes that manifest themselves as HLX, with X-ray luminosity $L_{x} > 10^{41} $erg/s as a function of projected distance. 
 We concentrate on the population of wandering HLXs in galaxies 
in two mass bins: $M_{*} > 10^{10} \msun$ and $10^{9} \msun < M_{*} < 10^{10} \msun$. We also show a prediction considering only luminosities $L_{x} > 10^{42} $erg/s.
Additionally, we explicitly display the contribution to the HLX population of seed IMBHs at the seed mass.
Currently, the largest sample of about $\sim 200$ HLXs has been assembled by \citet{Barrows2019} using (pretty massive) galaxies in the Sloan Digital Sky Survey (SDSS), cross-matched with Chandra X-ray catalogues. All of the galaxies in this study are at $z < 0.9$, i.e., at a significantly lower redshift than considered here. \citet{Barrows2019} find most of their sources to be broadly spread over a few tens of kpc. The peak of the observed HLX distribution is at $L_{x} \sim 4 \times 10^{42} $erg/s — for this reason, we also show \astrid \, predictions considering only HLX with $L_{x} >  10^{42}$ erg/s (dotted line).

\begin{figure*}
\centering
\includegraphics[width=2\columnwidth]{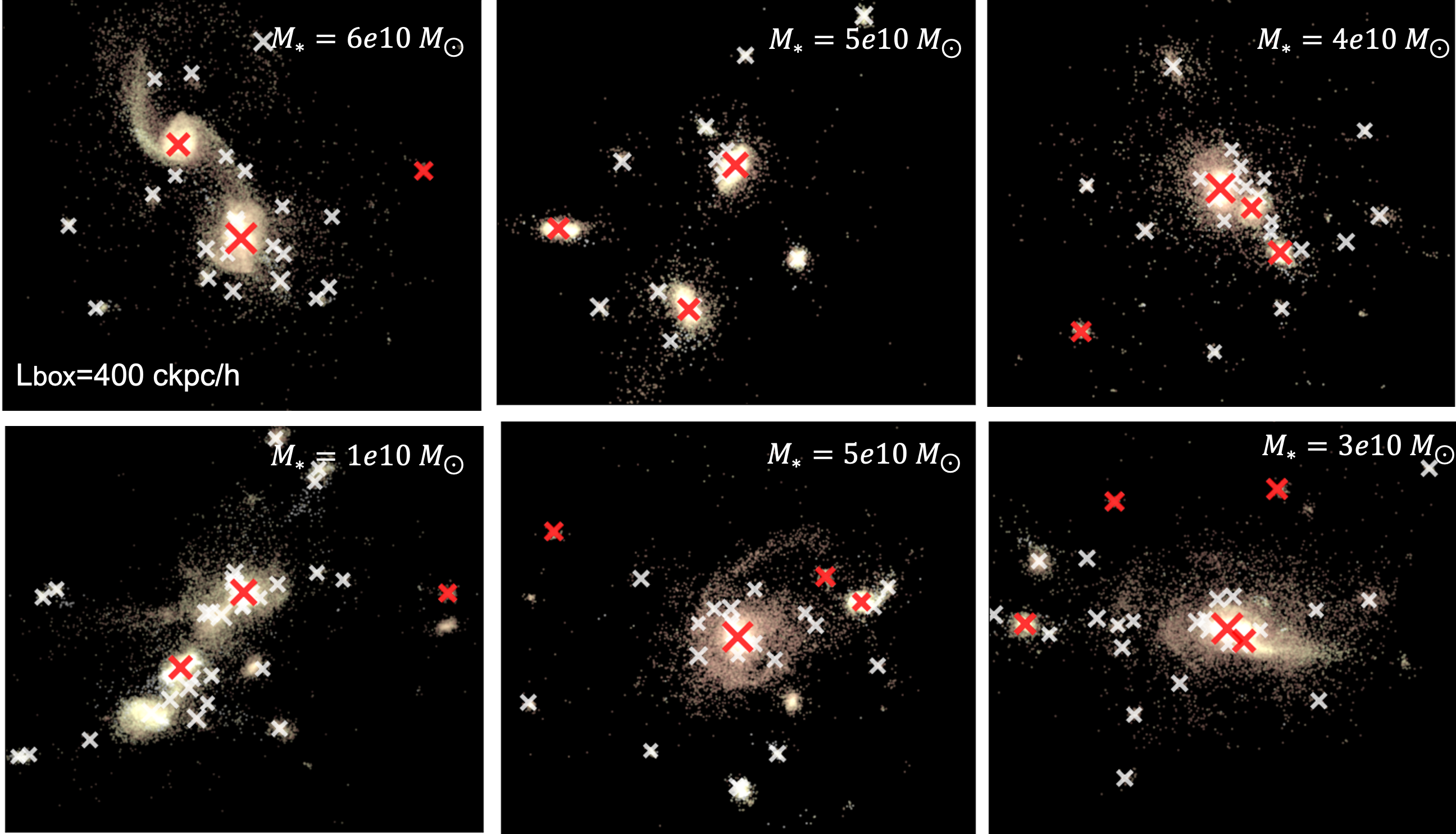}
  \caption{HLX sources (shown with red crosses) in 6 example galaxy environments, with stellar densities shown. At $z\sim 2$ HLX are both in galaxies, substructures around galaxies, and broadly distributed, consistent with the PDF shown in  Figure~\ref{fig:offsets1}.  The side length of these images is 400 ckpc/h.
  }
  \label{fig:HLXmap}
\end{figure*}

\astrid \, results show a broad and rather flat distribution of HLX sources as compared to the inferred distribution from \citet{Barrows2019}.
In the observed sample, the HLX distribution falls more quickly (at scales $> 20 $kpc) than the one predicted from the simulation 
(in both bins of galaxy stellar masses), even when considering the higher luminosity HLXs. 
We caution, however, that the comparison is only somewhat indicative, as we are making predictions at $z=2$ and a significant amount of infall is likely to take place.
As shown in Figure~\ref{fig:HLXmap} many of the systems 
are actually actively merging at $z\sim 2$.
The broadest, flattest distribution of HLXs is displayed in the high stellar mass bin due to IMBHs residing in satellite galaxies and significant substructures within and in the outskirts of the main galaxy (and likely to infall at later times). 
\citet{RicarteIMBH2021} perform a more detailed comparison (at $z=0.9$) between simulation and observations, finding broad agreement with the observations and also a broad distribution of offset HLXs. 

Here we find that virtually none of the seed IMBHs would have been detected in samples with average $L_x \sim 10^{42} $ erg/s.
At $L_x \sim 10^{41} $ erg/s IMBHs make a contribution, and
roughly 1 in 10 HLX is due to a seed IMBH. Seed IMBHs are somewhat more prominent at larger offsets compared to the rest of the wandering IMBHs. 

The right panel of Figure~\ref{fig:ULXoffsets} shows the relationship between $M_{\rm BH}$ and the host galaxy stellar mass $M*_{\rm host}$ for the entire population of wandering BHs which are also HLX sources in \astrid. 
Here we use the subhalo information in \astrid, and consider the wandering BHs as residing within subhalos. 
With the high-resolution of \astrid, we find that the majority of ``wandering'' seed IMBHs are actually associated with small dwarfs, with $M_*{\rm host} \lesssim 10^{7} \msun$, and the majority of these are embedded in the large groups. These host galaxies are expected to be hard to detect in any follow-up of HLX sources.
It is also striking that the largest numbers of HLXs are indeed IMBHs with $M_{\rm BH} \lesssim 10^{6} \msun$ associated with dwarf galaxies with $M_*{\rm host} \lesssim 10^{8} \msun$.
There is also a large number of IMBHs in large galaxies: these have likely been fully stripped of their host galaxies \citep[as discussed in][]{Chen2022} and
truly 'wandering' in massive host galaxies and accreting as they experience large density environments in the central regions \citep[][]{Weller_2022, Seepaul_2022}.

The inferred BH masses and detected host galaxies from \citet{Barrows2019} (red points in left panel of Fig.~\ref{fig:ULXoffsets}) are included in the HLX sample in \astrid \, but they are indeed unlikely to have probed the bulk of the population of seed IMBHs.

Finally, we also find that a relatively small number of HLX sources are comprised of massive BHs (in massive host galaxies)
and likely the remnants of major mergers, now offset and orbiting a central SMBH.
Simulations do not predict that there should be a clear relation between $M_{\rm BH}$ and $M_{*}$ for the HLX population (in particular those due to IMBHs). IMBH HLX sources sample a wide range of stellar mass for their host galaxies.

\section{GW signatures of IMBH inspiral and mergers}
\label{section:GW_signatures}

\begin{figure*}
\centering
\includegraphics[width=1\textwidth]{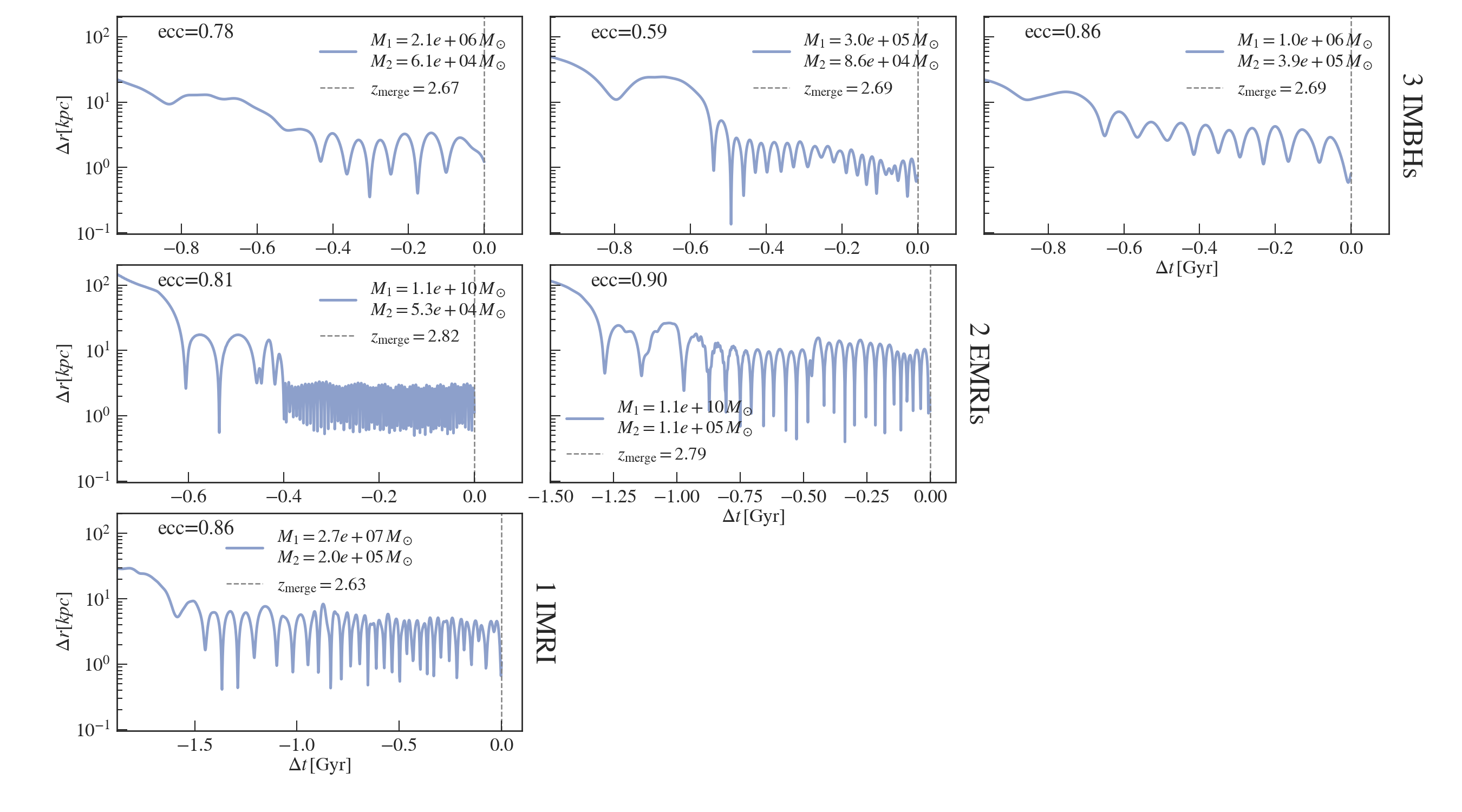}
  \caption{The inspiral of five pairs of IMBHs mergers. Top row: the three IMBH-IMBHs mergers in Group 3 (see Figure~\ref{fig:group40}). The two EMRIs in the middle row are also shown in Figure~\ref{fig:group21}. Bottom row: an IMBHs merger with a supermassive BH.
  The distance $\Delta r$ for the pair is plotted against time, where $t=0$ is the time of the merger. The inspiral-to-merger time can be read off the x-axis. 
  }
\label{fig:5inspirals} 
  \end{figure*}
By $z\sim 2$, in \astrid, IMBHs populate halos of all masses and 
they can be as numerous as several hundred in massive galaxies.
The largest fraction of the wandering IMBH population is composed of seed mass BHs that have been incorporated into larger and larger halos via successive, prior minor halo and galaxy mergers.
At high-z, the large population of IMBHs in halos is the result of inefficient dynamical friction and the inability to sink IMBHs to the center of the shallow potentials of high-z galaxies. 
This results in massive galaxies at $z \sim 2$ building up the large population of wandering IMBHs which have been stripped of their host galaxy and left orbiting in  halos.
In the previous section, we have seen that other IMBHs are infalling with their host galaxy, and finally, a large number of satellite galaxies and their IMBHs are orbiting in the outer regions of massive
galaxies (as seen also in the example in Section \ref{sec:IMBH_illustration}).
With these large numbers of IMBHs in the 
deeper potential, dynamical friction is more effective and IMBHs start to sink and merge more effectively at these $z\sim 2$.

In the previous Section, we have shown that HLX sources only reveal a small fraction of the IMBH population and even fewer of the most numerous seed IMBHs: as expected, most of the seeds that 'survive' to $z=2-3$ have not been accreting effectively and, hence, remain mostly hidden while wandering in galactic halos.

The large population of seed IMBH we find at $z=2-3$ is a result of what is often referred to as the "sinking problem" for IMBHs at high-z. Regardless of the specifics of any seed model dynamical friction (DF) is ineffective:
cosmologically simulated galaxies in the early universe with a variety of DF implementations have consistently shown that
massive seeds ($\sim 10^{5} \msun) $
do not sink to the center of high-z galaxies — no matter the implementation of DF used \citep{Pfister2019, Tremmel2018, Ricarte2019,  Ma2021, Chen2022}.
As discussed  in these works, at high-z it is highly unlikely for BH seeds to sink in the center of galaxies and hence merge. However, at some point,  BHs (even IMBHs) become anchored to the galaxy center,   as their host galactic centers themselves are better defined and dense enough to allow them to sink and possibly merge.
In particular, \citet{Tremmel2018a-CHANGA} showed that, indeed, as galaxies become massive enough with unambiguous massive central peaks
at intermediate redshifts ($z=2-4$), sinking, dynamical interactions, and mergers become more predominant.
This is exactly what \cite{Chen2022} found using \astrid, and showed that the BH mergers are extremely rare at high-z but BH merger rates increase ($\sim 1 $ per year) and peak by $z=2-3$. In general, these results show that the properties of the merging galaxies (mass ratio, central density) matter even more than the mass of the black hole when determining their orbital evolution. As discussed in \citet{Tremmel2018a-CHANGA}, for the most rapidly forming MBH binaries, it is the galaxy interaction that bring the black holes together, rather than dynamical friction acting on them individually.

Since there is a large time delay between the pairing of the BHs (by seeds or other black holes) and its coalescence, dictated by the efficiency of dynamical friction (and the hardening processes, see also below) seed IMBHs can merge in galaxies of all types, as in this lapse time that can be of the order of Gyrs, host galaxies undergo infall, mergers and strong evolution. Thus, coalescing seed IMBHs formed in halos at high redshifts can track all different environments.
We have seen that there is a rather large concentration of IMBHs within the central regions of galaxies — therefore, we expect that IMBHs will start merging more effectively and be detectable by LISA \citep[based on the delay timescale estimated in e.g.][]{Chen2022}. At  $z\sim 2$ BH mergers will directly probe the seed IMBH population in all sorts of host galaxies.

Between z=2 and z=3 there are a total of 826,000 BH mergers in \astrid, the overwhelming majority of them being between seed-seed IMBHs ($q\sim 1$), 100,000 with $q=10^{-1,-2}$, and $\sim 10000$ $q=10^{-2,-4} $ (IMRIs) and 150 BH mergers (EMRIs) with $q< 10^{-4}$. IMRIs and EMRI involve IMBHs which are in high-density regions around supermassive black holes in the centers of massive galaxies.

\subsection{Illustrative Examples: Inspiral and Mergers}
In Figure~\ref{fig:5inspirals} we show the
inspiral and orbital decay for the five seed
IMBH mergers (indicated with $t=0$) which we  highlighted in Section \ref{sec:IMBH_illustration}   (Figures~\ref{fig:group9}, \ref{fig:group40} and \ref{fig:group21}. The top three panels illustrate the inspiral of three examples of seed-seed
BHs ($q \sim 1$). 
We see that the timescale between pairing, inspiral, and merger in \astrid, as expected, approaches $\sim 1$ Gyr for these IMBH pairs.
Many of the seed IMBHS have not merged, but for these examples (and many of the other mergers close to these redshifts) the orbital eccentricities are typically high, between $\sim 0.6-0.9$ \citep{Chen2022}.  
If we consider the timescale associated with the hardening (both additional dynamical friction and loss cone scattering timescales) for these three mergers \citep[done in postprocessing in][]{Chen2022}, two of these mergers ($z_{\rm merge} \sim 2.6)$ are delayed to $z \sim 2$.
For one of the examples involving IMBHs with $M_{1}=10^{6} \msun$ and  $M_{2} = 4\times 10^{5} \msun$,  the hardening timescale exceeds the Hubble time.
 
In the middle row of Figure~\ref{fig:5inspirals} we show the inspiral of two EMRIs, with binary systems in which one component is a massive black hole of $10^{10} \msun$  (at the center of a massive galaxy) 
and the other is a seed IMBH of $\sim 10^{5} \msun$ (this is the example shown in Figure~\ref{fig:group9}). 
The small mass ratio leads to slower evolution of the binary. It is apparent (even at the scales resolved in \astrid) that the seed IMBH completes many tens of orbits: the first merger (shown on the right) with $q \sim 10^{-5}$ 
takes over $1.4$ Gyrs, while the second event (shown on the left), with $q\sim 5\times 10^{-4}$ results from the dynamical interaction of the three bodies (two IMBHs and an SMBH). 
The inspiral for the second event takes about 1/2 the time of the previous EMRI, and the merger occurs in about 0.5 Gyrs.
These EMRIs are rare, (only 2 out of $10^4$ mergers) and the large volume of \astrid~is crucial for capturing these remarkable events. 
In massive systems, the large population of sinking seed IMBHs and central SMBHs gives rise to EMRIs. 
In  massive galaxies IMBHs find themselves in a rich, dense stellar environment around galactic nuclei: this may lead to additional dynamical friction
(calculated in postprocessing) adding 1-2 Gyrs delay to the merger (whilst loss cone scatter remains subdominant). 
Even if these delays are accounted for, we would expect these EMRIs to produce merger events at $z \sim 1.8$ (for the lowest mass seed) and $z \sim 1.2$ for the higher mass of the two IMBHs.

The bottom row of Figure~\ref{fig:5inspirals} shows the example IMRI. 
This BH binary is orbiting and inspiralling a relatively massive satellite, which is itself infalling toward the central massive galaxy (as shown in Figure~\ref{fig:group21}).
The BH binary involves an SMBH with a mass of $2.7 \times 10^7 \msun$  and a seed of $2\times 10^{5} \msun$. 
This is the slowest inspiral, taking over $1.7$ Gyrs to merge. The merger in \astrid \, occurs at $z\sim 2.6$.
Post-processing this event and accounting for binary hardening processes imply delays longer than the Hubble time for this particular example. 

\subsection[LISA Events from IMBHs at redshift 2]{LISA Events from IMBHs at $z \sim 2$} 
\begin{figure*}
\centering
\vbox{
\hbox{
  \includegraphics[width=0.5\textwidth]{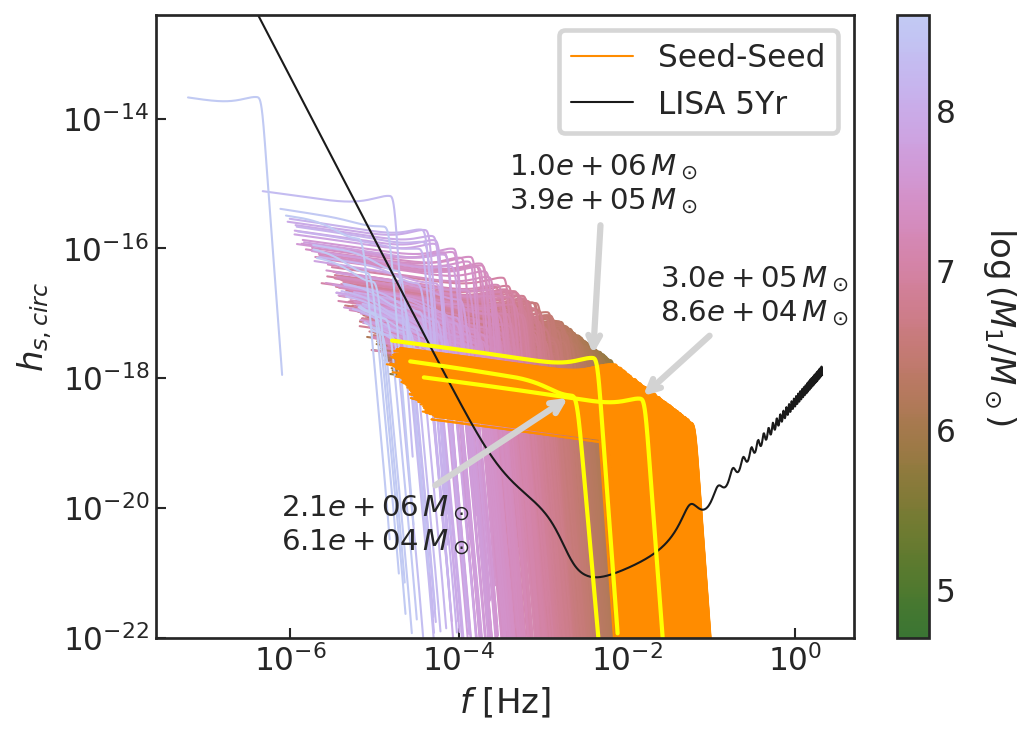}
  \includegraphics[width=0.5\textwidth]{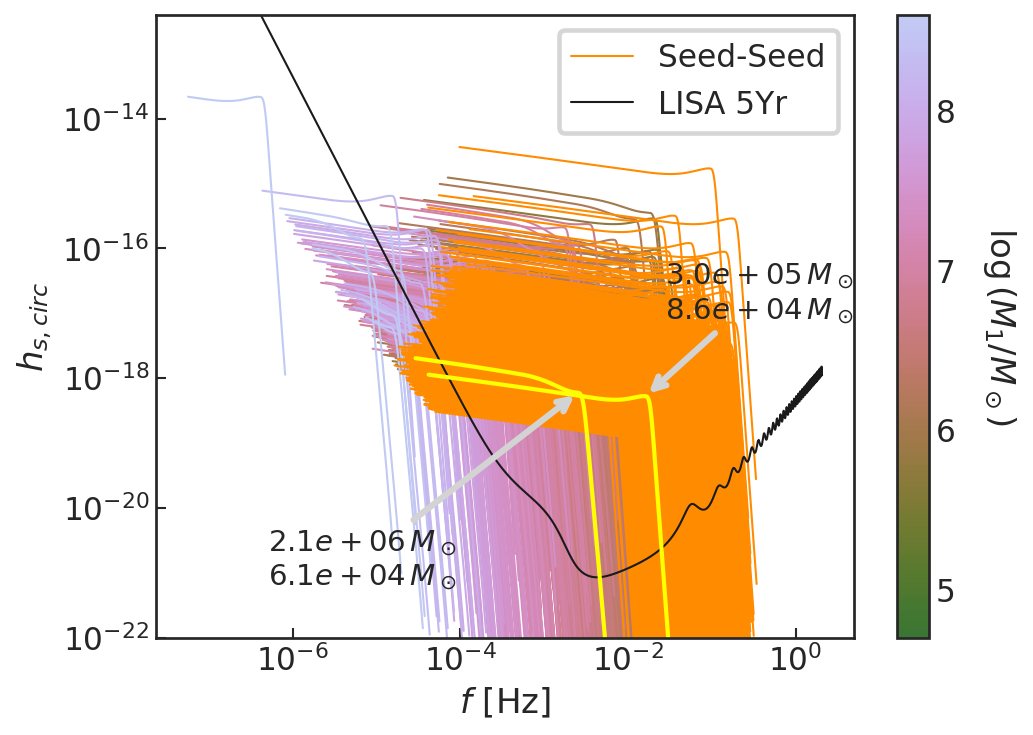}
  }
  \hbox{
  \includegraphics[width=0.5\textwidth]{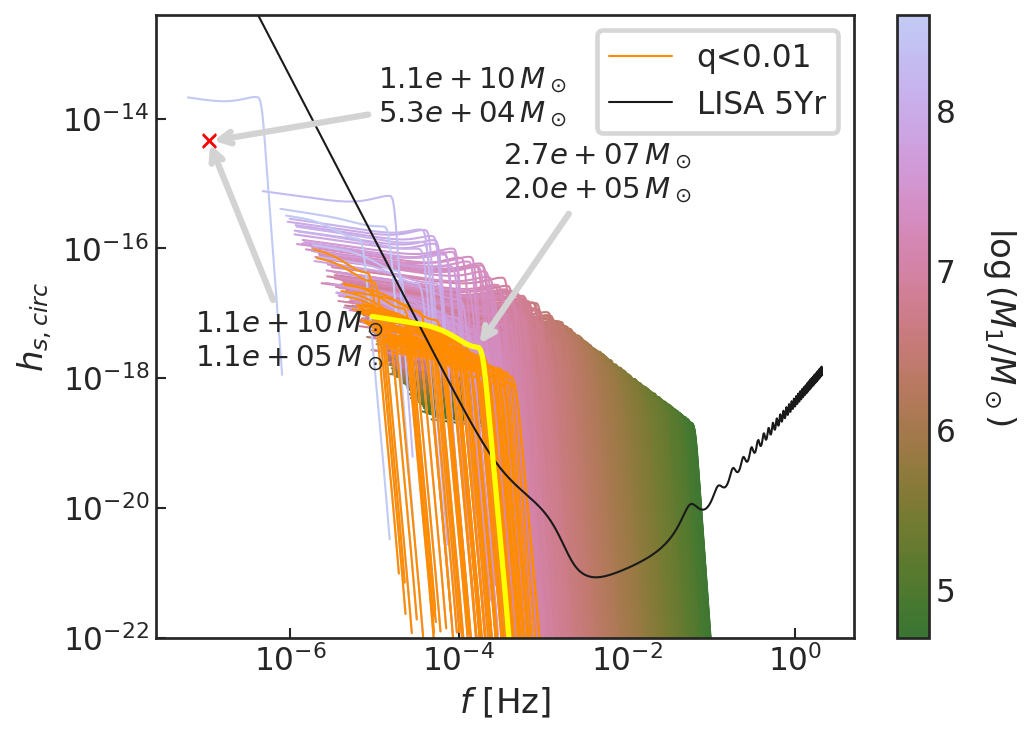}
  \includegraphics[width=0.5\textwidth]{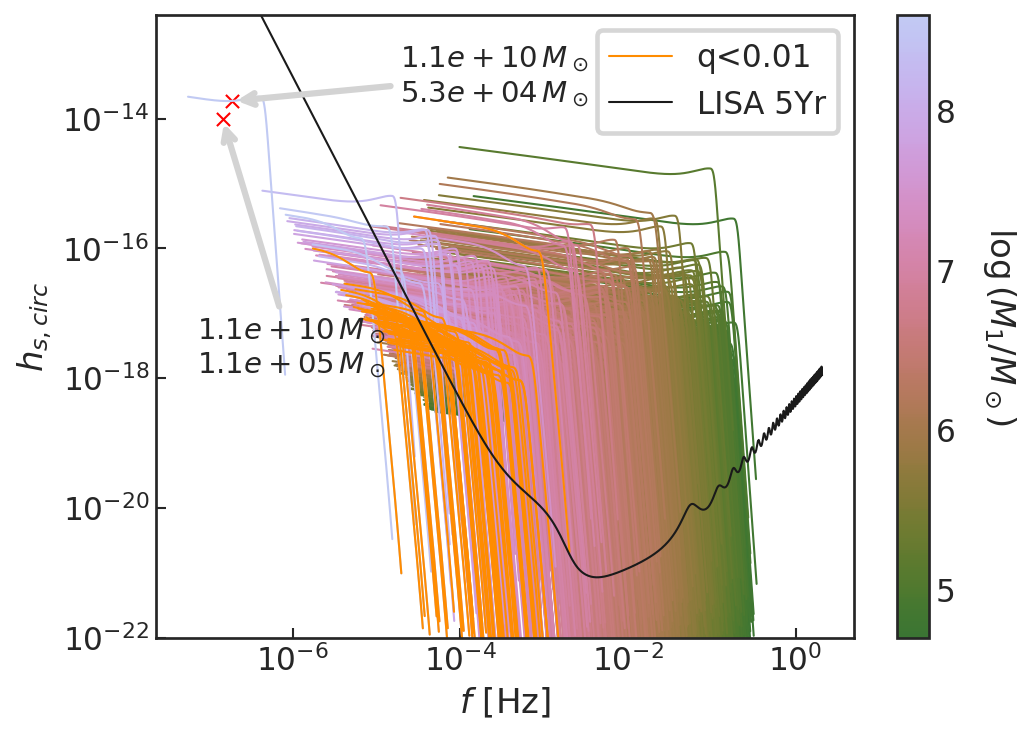}
  } 
  }
  \caption{The strain amplitude as a function frequency for the IMBHs inspirals and mergers in \astrid. A subset of the events is shown (pink to green) for events involving massive BHs to IMBHs. The yellow colors indicate events involving IMBHs (with $q\sim 1$) in the top row. In the bottom row, we display events with $q < 0.01$ involving one IMBH and a massive BH. The plots in the right column show the effect of additional (post-processed) hardening and loss cone scattering which delays the mergers toward lower redshifts.}   
  \label{fig:mergerstrain} 
\end{figure*}

In the next decade, LISA will observe the coalescence of massive black hole binaries in the mass range $3 \times 10^3 \msun < M_{BH} < 10^7 \msun$ \citep{Cosmo_LISA2022, Astro_LISA2022}.
Figure~\ref{fig:mergerstrain} shows the LISA sensitivity (black line, LISA sensitivity \citealt{Robson2019}) with a detailed prediction of the expected and potential sources from \astrid. In particular, we highlight in orange the events crossing the LISA band that involve IMBH seeds. \citep{Chen2022} already discussed that the great majority of LISA events at $z=2-3$ are likely to involve seed black holes. LISA detection of $q\sim 0.1-1$ (shown in the top panels) with a strain $h_{c} \sim 10^{-18}$ from these redshifts could in principle be capable of providing constraints on the seed BH mass function. These event rates amount to about $\sim 1$/yr. In practice any LISA detection, at $z\sim 2-3$ will most likely be due to the inspiral of seed IMBHs.

Massive  galaxies, instead, where the number of IMBHs that can be packed in the centre of a galaxy can be as high as 10-100 host the formation of MBH-IMBH pairs  These EMRI/IMRI-like systems can form at relatively low redshift.
Detecting IMRIs and EMRIs such as the example discussed
here will require next-generation GW detectors in the proposed milliHz-$\mu$Hz band \citep{Sesana_2021}.

In Figure~\ref{fig:mergerstrain} we show the detailed  IMBHs events in the strain-frequency plane, together with the LISA sensitivity curve. The left column shows the predictions for \astrid, while on the right all the events have been post-processed to account for the additional effects of hardening and loss cone scattering. In the top panel, the orange lines show all the seed IMBHs ($q\sim 1$) events filling a large region of the parameter space of LISA detections. In the bottom panels, instead, the orange highlights all the $q< 0.1$ events, EMRIs, and IMRIs, including those that lie outside the LISA sensitivity curve and would require future $\mu$ Hz space-based interferometers \citep{Sesana_2021}, which  would fill in the rather large gap in frequency between LISA and Pulsar Time Arrays (PTAs). The specific examples shown in Figure\ref{fig:mergerstrain} and illustrations in Section~\ref{sec:IMBH_illustration} are highlighted in yellow for reference.

If indeed a population of seed IMBHs exists, LISA events at $z=1-3$ will be dominated by these BHs that are expected to 'survive' and accumulate in large galaxies and finally undergo mergers at these redshifts.
This, in turn, implies that GW observations with LISA will likely provide the most stringent constraints on IMBH populations, which provide the dominant event rates.



In addition to IMBH-IMBH inspirals, we also  find intermediate mass-ratio inspirals (IMRIs) with a mass ratio in the range $10^{-2}$ to $10^{-3}$.

Within the class of compact-object binary systems, the ubiquity of SMBHs residing in galactic nuclei makes binaries with a mass ratio $q \sim 10^{-3,-2}$ a prime target for low-frequency space-based GW observatories. These systems we have already referred to as EMRIs – binary systems in which one component is substantially
more massive than the other – hold the exciting promise of providing unrivaled tests of Einstein’s GR \citep{Cosmo_LISA2022}. They also hold the promise to improve our understanding of the properties of galactic nuclei and to enable precision measurements of the properties of SMBHs out to high redshift

The small secondary completes many tens of thousands of orbits,  speeding in the strong gravitational field of the heavy primary. Furthermore, the binary tends not to have completely circularized, resulting in orbits with a very rich structure. The resulting
GWs carry with them exquisite information about the spacetime of the binary and the environment in which the IMRI lives. Regarding the latter, there remains a great deal of uncertainty about the nature of the stellar environment around galactic nuclei.
In addition to EMRIs, searches will target intermediate mass-ratio inspirals with a mass-ratio in the range $ 10^{-2}$ to $10^{-3}$.
In order to extract the maximum science gain from IMRI observations
(i) high signal-to-noise ratios will be needed, and (ii) a sufficient number of sources to draw statistically significant conclusions.



\section{Summary \& Discussion}
\label{section6:Conclusion}

The \astrid \, simulation predicts the existence of an
extensive wandering IMBH population at cosmic noon. IMBHs
are mostly composed of IMBH seeds formed at high-z and incorporated into increasingly large halos via successive mergers.
At $M_{\rm BH} \lesssim 10^6 \msun$ wanderers greatly outnumber central black holes. 
Over half the galaxies with $M_* \sim 10^9 \sun$ have offset IMBHs and for $M_* > 10^{10} \msun$ they number in the tens to several hundred. 
However, only a small fraction of this extensive population is likely to be revealed as HLX sources. 
The relative high densities of IMBHs in these dynamic environment leads to $\sim 800,000$ IMBH mergers in the \astrid \, volume at $z=2-3$. Many of these are IMBHs-IMBHs paired in dwarf galaxy satellites in the outskirts of massive groups. There are also a few thousand events involving massive black holes and IMBHs (IMRIs) and a few hundred mergers of supermassive black holes with IMBHs (EMRIs).
The extensive population of wandering IMBHs, and remnants of IMBHs seeds stand to be revealed most prominently via their GW signatures at cosmic noon.

The relatively short period between $z=2$ and $z=3$, referred to as cosmic noon, corresponds to the time when galaxies formed
about half of their current stellar mass (e.g., \citealt{madau14}). This epoch is also expected to correspond to the peak of BH merger activity \citep{VolonteriHorizon2020, Chen2021}.
\cite{Chen2021} found that this is indeed the case for \astrid~  even though the realistic dynamical friction modeling used in that study leads to delays between galaxy mergers and mergers of their black holes by a few hundred Myr on average.
Because of the rapid growth in activity from earlier redshifts to this peak, cosmic noon was likely the time when the first IMBHs will be seen, both by LISA and through their X-ray emission. For
example, at redshifts $z>6$ the number of mergers per year is expected to be at least 50 times less than at $z\sim 2$ (\citealt{Chen2021}).
It has already been shown in simulations (\citealt{Ma2021}) that at these earlier, reionization-era redshifts there has not been enough time for the majority of BH seeds to merge or grow significantly. \astrid,
with its large volume, has enabled us to follow the situation and number of seeds remaining down to redshifts where IMBHs will be detectable due to their EM emission.

Our findings for the IMBHs population in \astrid \, are consistent with those of \citet{RicarteIMBH2021, RicarteHLX2021} who used ROMULUS25 simulations. Here we are able to extend these types of studies to a larger volume simulation (\astrid \, is a factor $\sim 10^{3}$ larger volume
than most previous studies, allowing us to study directly the population of IMBHs in larger halos, where the highest IMBH densities are reached) and their interaction with their central SMBHs. \astrid \, also explicitly models seed BHs at masses $< 10^{6} \msun$ allowing us to directly probe the regime of IMBHs.
The large volume, together with the smaller seed mass than previous simulations, are crucial for predicting GW signatures in the LISA regime. However, \astrid thus far reaches only $z \sim 2$ and we cannot make direct predictions for the local population of IMBHs and associated EM signatures of this faint population \citep[for this, see][]{RicarteHLX2021}. Another important limitation of this study is our mass resolution, which requires that our BH dynamics be governed by a dynamical mass, $M_{\rm dyn} \sim M_{\rm DM} >M_{\rm BH}$. This choice is necessary to avoid numerical heating of BHs due to much larger background particles, but it also means that dynamical friction may be overestimated for the smallest black holes in the simulation. Therefore, the population of wandering BHs predicted in this work should be considered a conservative estimate. In reality, wandering BHs may be even more dominant at the lowest masses than what we predict in this work.

X-ray studies of HLXs at $z=2.4$ have been attempted \citep{Mezcua2018}, with some tentative detections, although at lower redshifts ($z=0.9$, see, e.g., \citealt{Barrows2019}) the
observational situation is firmer. Our study sheds light on the nature of IMBHs at cosmic noon, with our main conclusion being that about 1 in 10 of these sources are likely seed BHs.
We have seen that there is a significant overlap between the seed population and BHs with $L_{X}> 10^{41}$ erg/s, so X-ray observations will allow direct constraints to be placed on IMBH seeds.
A caveat here is that most simulations, including \astrid, do tend to overproduce the faint end of the AGN luminosity functions \citep{Habouzit2022}. Our luminosities for IMBHs rely on a subgrid Bondi-like accretion rate which is likely an upper limit.

There are many models for massive seed formation, ranging from runaway stellar collisions in high-redshift nuclear star clusters (e.g., \citealt{HB_2008, Komossa_Merritt_2008, Fragione_2018a}), massive metal-free stars (e.g., \citealt{pelupessy07}) to direct collapse from gas clouds (e.g., \citealt{BrommLoeb2003, Lodato_2006}). With \astrid, we have seen that much of the information from seed formation may be preserved down to at least to this $z\sim 2 $ epoch, and will be accessible to observations.
Such large populations of IMBHs 
should also lead to a significant number of IMRIs and EMRIs (IMBHs and stellar companions, not modeled in \astrid) to be uncovered by the planned decihertz  
gravitation wave observatories \citep[e.g.][]{ArcaSedda_deci2020}.


\section*{Acknowledgements}
\astrid \, is carried out on the Frontera facility at the Texas Advanced Computing Center. The simulation is named in honour of Astrid Lindgren, the author of Pippi Longstocking. SB was supported by NASA grant ATP-80NSSC22K1897.
TDM and RACC acknowledge funding from the NSF AI Institute: Physics of the Future, NSF PHY-2020295, NASA ATP NNX17AK56G, and NASA ATP 80NSSC18K101. TDM acknowledges additional support from NASA ATP 19-ATP19-0084, and NASA ATP 80NSSC20K0519, 
MT is supported by an NSF Astronomy and Astrophysics Postdoctoral Fellowship under the award AST-2001810. 
FP acknowledges support from a Clay Fellowship administered by the Smithsonian Astrophysical Observatory. This work was also supported by the Black Hole Initiative at Harvard University, funded by grants from the John Templeton Foundation and the Gordon and Betty Moore Foundation.

\section*{Data Availability}
The code to reproduce the simulation is available at \url{https://github.com/MP-Gadget/MP-Gadget} in the \texttt{asterix} branch. 
Part of the \astrid snapshots are available at \url{https://astrid-portal.psc.edu}.
The complete BH catalogues at $z>2$ will be available online shortly.

\bibliographystyle{mnras}
\bibliography{main.bib}

\end{document}